\documentclass[%
 preprint,
 amsmath,amssymb,
 aps,
]{revtex4-2}

\usepackage{tikz}
\usepackage{lineno,hyperref}
\usepackage{epsfig}
\usepackage{amssymb}
\usepackage{mathtools}
\usepackage{booktabs} 
\usepackage{enumitem}
\usepackage{graphicx}
\usepackage{subcaption}
\usepackage{pgfplots}
\modulolinenumbers[5]

\usepackage{dcolumn}
\usepackage{bm}

\DeclareRobustCommand\full  {\tikz[baseline=-0.6ex]\draw[thick] (0,0)--(0.5,0);}

\DeclareRobustCommand\dashed{\tikz[baseline=-0.6ex]\draw[thick,dashed] (0,0)--(0.54,0);}

\begin{document}


\title{Characterizing 1D Inertial Particle Clustering}

\author{Daniel Odens Mora}
\affiliation{
Univ Grenoble Alpes, CNRS, Grenoble-INP, LEGI, F-38000, Grenoble, France
}%
 \altaffiliation[Also at ]{Univ Grenoble Alpes, CNRS, Grenoble-INP, LEGI, F-38000, Grenoble, France}
\author{Alberto Aliseda}%

\affiliation{%
 Department of Mechanical Engineering, University of Washington, Seattle, Washington 98195-2600, USA
}%

\author{Alain Cartellier}
\author{Martin Obligado}
 \email{Martin.Obligado@univ-grenoble-alpes.fr}

\affiliation{
Univ Grenoble Alpes, CNRS, Grenoble-INP, LEGI, F-38000, Grenoble, France
}%

\date{\today}

\begin{abstract}
Preferential concentration is a common phenomenon found in turbulent flows seeded with inertial particles. Although it has been studied extensively, there are still many open questions about its fundamental physics. These gaps hinder the reconciliation of different experimental and numerical studies into a single coherent quantitative view, which is needed to enable accurate high resolution modeling.
	
	This work examines the influence of the measuring technique, and in particular of the dimensionality of the measurement (1D line, 2D planes or 3D volumes) on  the characterization of cluster properties (a consequence of preferential concentration), and proposes an approach to disentangle the cluster-characterizing results from random contributions that could contaminate the cluster statistical analysis.
	
	First, we studied this effect by projecting 2D and 3D data snapshots containing clustering onto a 1D axis. The objective was to simulate 1D sensors (widely used experimentally) with different sensing lengths. 
	
	These projected records were  analysed by unidimensional Vorono\"{i} tessellations. Our results demonstrate that average mean clustering properties could be retrieved, if the measuring window is equal or larger than the Kolmogorov length scale ($\eta$), and smaller than about ten percent of the integral length scale of the turbulence $\mathcal{L}$. These observations are consistent with 2D and 3D data taken under similar experimental conditions. This agreement validates our projection approach as it adequately captures the behavior of a 1D probe immersed in a 2D, or a 3D flow.
	
	Additionally, we found that in 1D the raw probability density function (PDF) of Vorono\"i cells does not provide error-free information on the clusters size or local concentration. We propose a methodology to correct for this measurement bias, based on the histograms of the number of particles within a cluster. The analysis of these histograms helps to explain the power-law behavior previously observed in the clusters size PDFs in 2D and 3D data. Moreover, the histogram analysis shows that such power law is not very robust in 1D, as it expands less than one decade.

	Finally, we show that to condition the statistics with the number of particles inside each cluster also allows to discern between turbulence-driven clustering and particle concentration fluctuations due to randomness. Our test then complements the classical cluster identification algorithm.
	
\end{abstract}

\maketitle


\section{Introduction}

Turbulent flows laden with inertial particles constitute an active research area within multiphase fluid mechanics due to their potential applications in fields such as: planetary formation, pollutant dispersion, and cloud formation \cite{shaw2003particle,larsen2018fine}. Hence, experiments are not only useful to validate the multiple numerical approaches, but also to disentangle their underlying physics, and finally reach comprehensive understanding. 

Several methods are available to characterize particle-clusters, with Vorono\"{i} tessellations \cite{Ferenc2007,huck2018role,yuan2018three,chouippe2019influence} becoming increasingly popular in both experimental studies employing visualization techniques, e.g., particle tracking velocimetry (PTV)  \cite{sumbekova2017preferential,huck2018role} and numerical simulations \cite{baker2017coherent,monchaux2017settling}. 

Despite its current popularity, it was early recognized \cite{Monchaux2012} that there could be pitfalls in the Vorono\"{i} method that could affect its results. The open questions are: to what extent physics from a 3D particle spatial distribution measured by 2D or 1D techniques can be captured by the Vorono\"{i} method, and how the measurement configuration (1D, 2D, line diameter, plane thickness, etc.) could impact its results. The study of Monchaux \cite{Monchaux2012} was the first to survey these possible biases in context of particle laden-flows. By projecting a 3D numerical dataset onto a 2D plane, they studied the effect of the laser sheet thickness ($L_{th}$ in their notation) on the resulting 2D Vorono\"{i} statistics. The study concludes that 2D Vorono\"{i} analysis is
reliable for the most common values of laser sheet thicknesses ($L_{th} \in [2\eta-6\eta] $) used in 2D imaging experiments, and very robust under random thinning, which simulates the effect of missed particles. However, care is needed when comparing 2D Vorono\"{i} statistics from different data sets obtained under different experimental or optical conditions, as sub-sampling results heavily depend on the particle seeding and the turbulence scales.

On the other hand, a recent study \cite{Mora2018} reports that a 1D Vorono\"{i} analysis (referred also as 1DVOA from now on) performed on a record taken by an optical probe via phase detection \cite{hong2004characterization}, is unable to capture preferential concentration in experimental conditions under which clustering is observed using 2D Vorono\"{i} analysis \cite{sumbekova2017preferential}. Thus, 1D and 2D Vorono\"{i} analyses could yield contradictory results regarding the presence of preferential concentration within the flow.

The study of Mora et al.\cite{Mora2018} suggests that the probe's measuring volume (a region in space where the instrument can detect particles transiting) has an important impact on this problem. This study reveals that below a certain threshold of the measuring volume, the 1D Vorono\"{i} method is unable to capture the spatial correlations between the recorded particles. This observation, however, is strongly affected by the particle concentration, as it was proven in 2D projections \cite{Monchaux2012}, and therefore, it cannot be attributed to insufficient resolution of the measuring instrument. Indeed, it is found that while an optical probe (with a measuring length 10 times smaller than the Kolmogorov length scale of the flow) was unable to retrieve clustering, it did recover the local liquid fraction, the particle size, and the particle velocity distributions consistent with previous measurements. 

\textcolor{black}{Biases involving 1D measurements have also been found in other methods to determine the existence of preferential concentration. For instance, when using 1D pair correlation functions, it has been found that excluded volume effects (due to the finite size of the droplets) could accumulate and had a combined effect across a range of scales, biasing the results obtained by this method \cite{kostinski2001scale,Shaw2002}. There is, however, a fundamental difference between 1D pair correlation functions and 1D Vorono\"{i} tessellations. The former method subdivides the dataset domain into segments of a certain scale, whereas, the latter method conducts analysis that is considered to be scale free. Excluded volume effects do not accumulate in 1D Vorono\"{i} tessellations given that 1D Vorono\"{i} cells are computed from their immediate neighbours.
	Conversely, the finite particle size impacts the value of the criterion used to assess the presence of clustering; clustering is present if the standard deviation of the Voronoi cell size $\sigma_\mathcal{V}$ coming from the particle dataset is larger than the respective one coming from the equivalent random Poisson distribution (RPP).
	Uhlmann has reported \cite{uhlmann2014sedimentation,uhlmann2017clustering,uhlmann2020voronoi} via 3D Vorono\"{i} tessellations that, for spherical particles with diameters up to $5\eta$, the respective magnitude of RPP standard deviation is reduced by 10$\%$. These observations imply that the nature of the pitfalls found in pair correlation functions, and 1D Vorono\"{i} tessellations, although related, are distinct. } 

Given the available technology, and experimental apparatuses, one could question the use of a `1D' technique to characterize a 3D phenomenon. Still, these complementary experimental techniques, for example, phase detection optical probes \cite{hong2004characterization}, or phase doppler interferometry (PDI) \cite{bachalo1984phase}, have been proven useful, as they provide additional information not so easily available from 2D/3D measurements. Moreover, when identifying the actual droplet spatial distribution within a cloud \cite{shaw2003particle,Jaczewski2005,larsen2018fine} or probing pollutant dispersion within cities \cite{Britter2003,Sabatino2013}, in situ measurements are usually acquired by 1D eulerian instruments. 

Hence, it is worth examining whether the 1D Vorono\"i statistics have any extra inherent biases when used to quantify preferential concentration, i.e., when the approach to analyze 2D or 3D data is extended to 1D records. The latter approach usually consists on building the Vorono\"{i} sizes probability density functions (PDFs), and then draw conclusions based on the particle local concentration \cite{Obligado2014,huck2018role,sumbekova2017preferential,uhlmann2014sedimentation,baker2017coherent,uhlmann2017clustering}, which is easily available via Vorono\"{i} tessellations.
\textcolor{black}{The presence of these biases is not restricted to 1D probes. Analogue problems could arise in 2D and 3D measurements. The 2D Voronoi analysis, however, has been shown to be very robust for different laser sheet thicknesses and concentrations \cite{Monchaux2012}.}

In this work, we study the impact of the instrument measuring window size (MWS) on 1DVOA statistics conducted on `projected' records from numerical and experimental data sets. Then, different MWSs `emulate' eulerian measurements taken with probes of different sizes in the same particle-laden flow. The later analysis allowed us to provide a range of MWSs (or probe sizes) for which evidence of preferential concentration could be consistently retrieved, under similar experimental values of the Taylor-based Reynolds number ($Re_\lambda$), and global liquid fraction ($\phi_v$).

According to our analyses, this range's lower bound (labeled MWS$_\dagger$) is of the same order of the Kolmogorov lengthscale ($\eta$). In other words, our analyses suggest that to retrieve clustering for similar values of $Re_\lambda$, and $\phi_v$ \textcolor{black}{one should measure with a probe with a window size of order $\eta$, i.e., MWS$_\dagger=O(\eta)$.
On the other hand, we observe that for the analyzed experimental data and DNS  (Direct Numerical Simulation) dataset this range's upper bound (labeled MWS$_\star$), after which evidence of clustering becomes weaker, seems to be located at measuring window size close to a tenth of the integral length scale.  MWS$_\star\sim \mathcal{L}/10$ where $\mathcal{L}$ is the integral length scale of the carrier phase.}

Furthermore, to characterize the clusters properties, we looked for biases in the probability density functions (PDFs) of cluster local concentration, and linear size. Our data strongly suggest that the method developed to analyze preferential concentration in 2D \cite{Monchaux2010} cannot be (in general) directly extended to 1D Vorono\"{i} tessellations analyses. Such direct extension, under some experimental conditions, is not conclusive regarding the presence of clustering within the flow, and undermines the capacity of 1DVOA to characterize the phenomenon in 1D unambiguosly. The origin of this limitation could be attributed to the loss of information intrinsic to the 1D measuring technique, and it is supported by the weaker correlation seen among the recorded particles positions.

To further explore these biases, we developed a theoretical model to compute the PDFs of cluster sizes coming from particle positions spatially distributed according to a random Poisson process (RPP). The model underlying principles suggest that conditioning the clusters PDF by the number of particles within clusters could aid to properly find and characterize `turbulence' driven clusters. The method, therefore, allows to disentangle randomness (as high density regions are also present within a RPP distribution due to random fluctuations of the local concentration) from turbulence (where clusters are controlled by the topology of the turbulent flow) in a 1D signal.

\section{The experimental setup and methods}
\subsection{Experimental setup and numerical database}

\begin{figure}
		\centering
		\begin{subfigure}[b]{\textwidth}
	\includegraphics[scale=0.9]{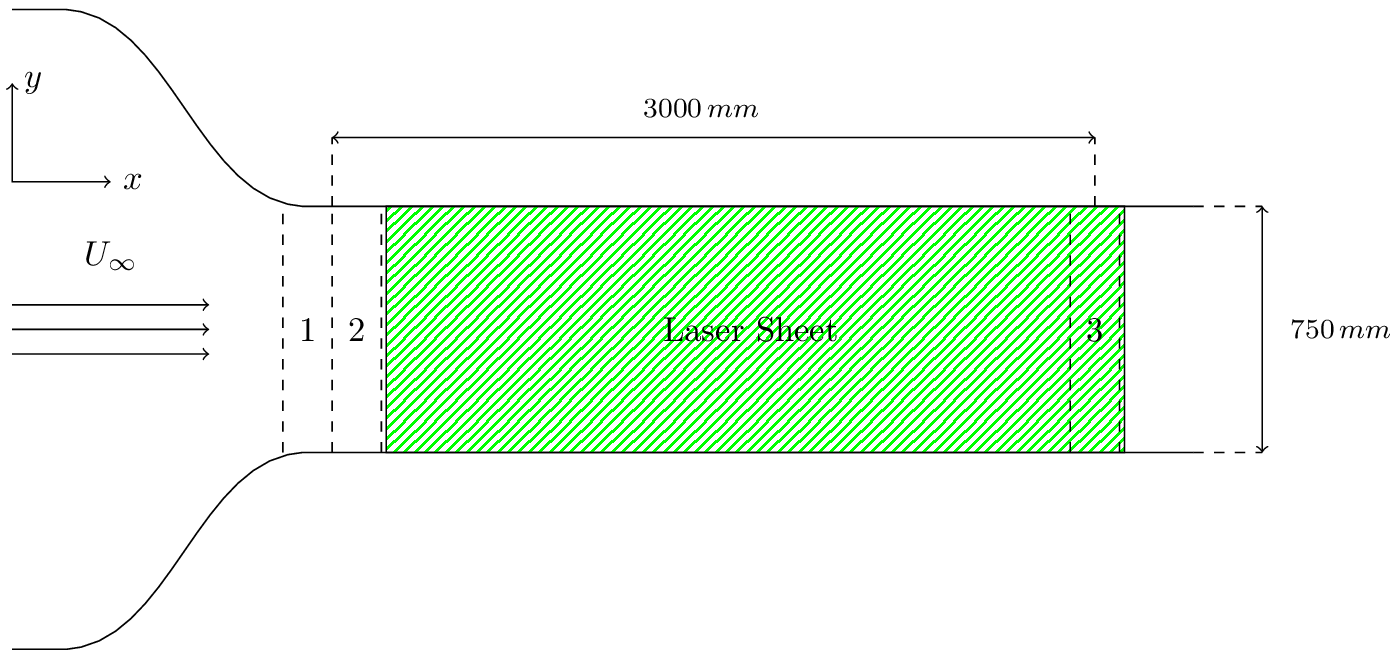}
	\caption{\label{fig:WT_sk}}
		\end{subfigure}

		\begin{subfigure}[b]{0.48\textwidth}
			\includegraphics[scale=.4]{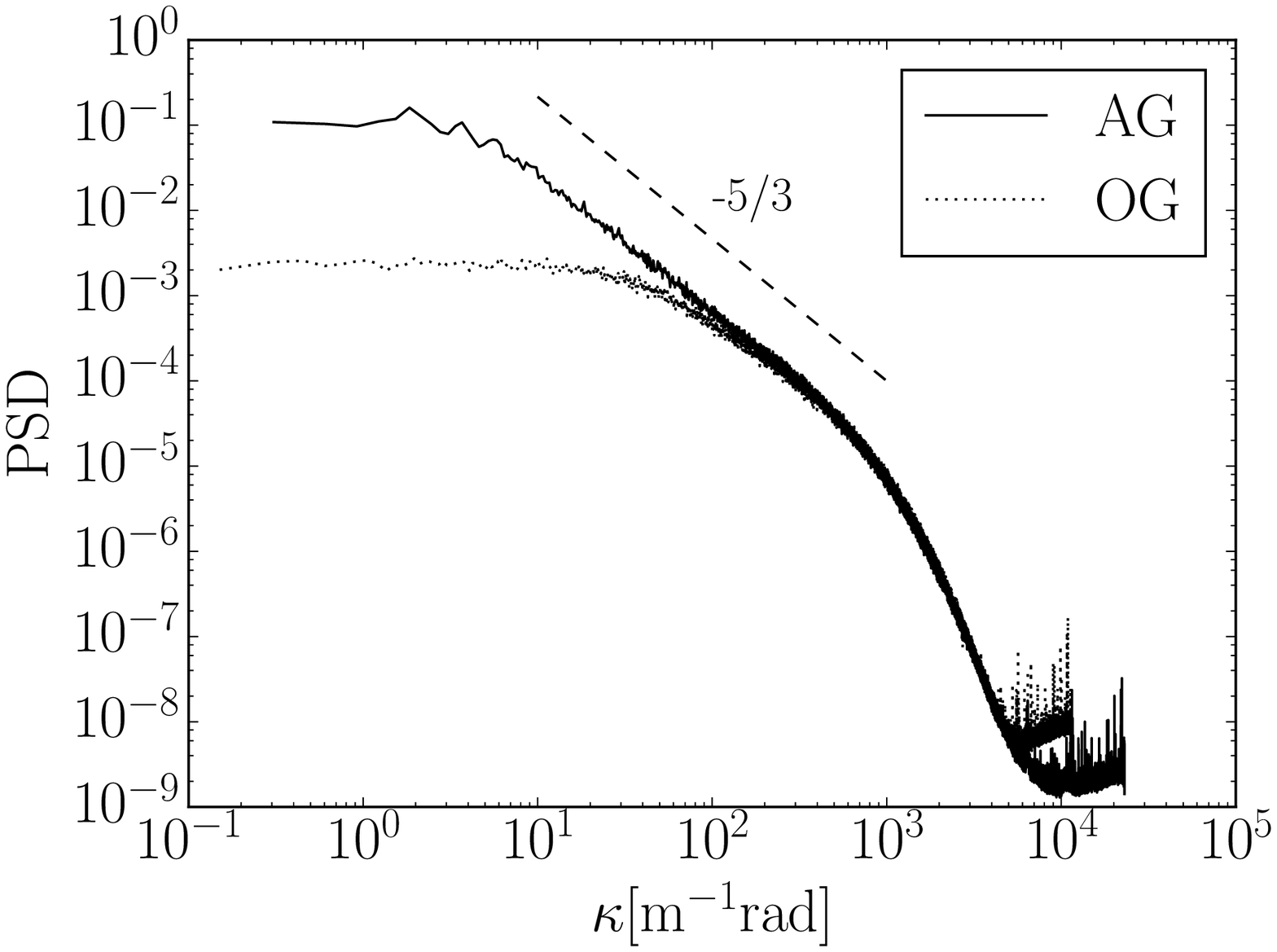}
			\caption{\label{fig:spk}}
		\end{subfigure}
		\begin{subfigure}[b]{0.48\textwidth}
			\includegraphics[scale=.4]{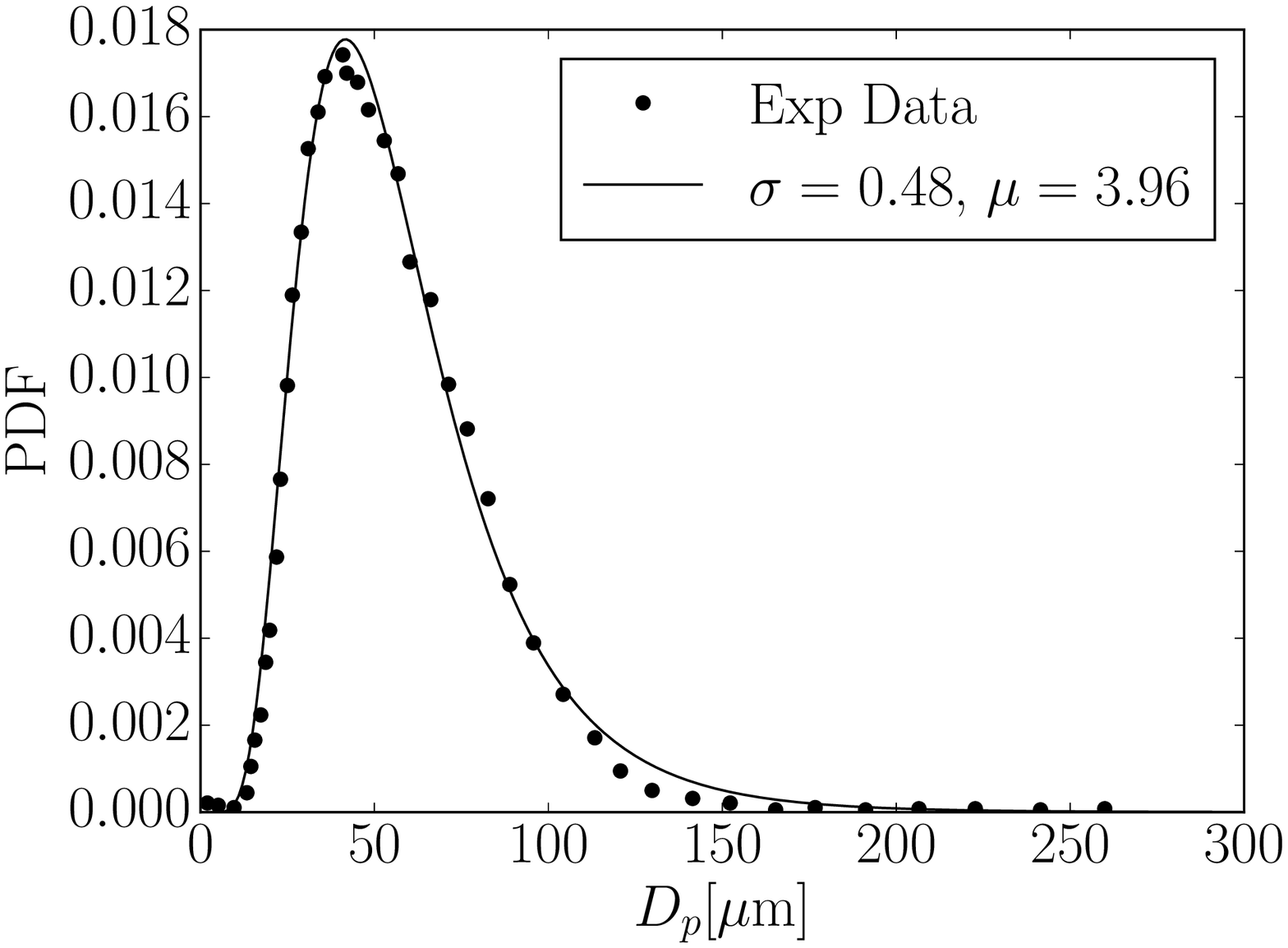}
				\caption{\label{fig:spray_pdf}}
	
		\end{subfigure}
	
		\caption{ a) Sketch of the wind tunnel experimental setup.  1, 2, and 3 refer to the locations of the active grid, the injection rack, and the measurement region, respectively. The measuring region downstream distance was taken from the beginning of the injector rack. The shaded region illustrates the extend of the laser sheet. The transverse square cross-section has dimensions of 750 $\times$ 750 mm$^2$.  b) Velocity power spectral density for the active grid (AG) $Re_\lambda \approx 250 $, and for the open grid (OG) $Re_\lambda \approx 30 $. Both spectra were obtained via hot-wire anemometry. c) Droplet Diameter $D_p$ distribution. The symbol ($\circ$) refers to data from \cite{sumbekova2017preferential}, and the (\full) line refers to a log fit (parameters shown in the plot legend) }

\end{figure}

The experiment was conducted in the close-circuit wind tunnel `\textit{Lespinard}' that has been extensively used to study particle clustering under turbulent conditions \cite{Monchaux2010,sumbekova2017preferential}. A sketch of the experimental setup is depicted in figure \ref{fig:WT_sk}.

Turbulence was produced by means of an active grid \cite{Mydlarski2017} operated in triple random or open mode \cite{mora2019energy}. At the measuring station (region $\#$3 in fig~\ref{fig:WT_sk}), the unladen turbulence has been experimentally found to be very close to a statistically homogeneous isotropic state (figure \ref{fig:spk}) under similar conditions of $Re_\lambda$, and $\eta$ \cite{Sumbekova2016a} to those found in the present work. 

Downstream of the grids, a rack of 36 spray nozzles generated inertial water droplets with a polydisperse diameter distribution. This polydispersity (see figure \ref{fig:spray_pdf}) has been previously quantified via phase doppler interferometry \cite{sumbekova2017preferential}. 

At the measuring station, a high speed camera collected 4500 images of the light scattered by the droplets from a 1-mm thick laser plane. The images, with an area of (120$\times$100 mm$^2$) and collected to be statistical independent realizations, were post-processed to identify the location of the droplet centers (more details about the experimental setup can be found in \cite{obligado2019study}). 

Along with experimental data, 30 numerical snapshots from a DNS data base \cite{benzi2017turbase} (\url{https://turbase.cineca.it/init/routes/#/logging/view_dataset/3/tabmet }) were surveyed.
Each file contained the trajectories of 1280 particles positions integrated over 3300 time steps.

These snapshots contained inertial point particles with a Stokes number close to unity; St$_\eta=\tau_p/\tau_\eta \approx1$ where $\tau_p= \rho_pD^2_p/18\mu$ is the particle relaxation time,  $\tau_\eta=\sqrt{\nu/\varepsilon}$ is the Kolmogorov timescale, $D_p$ is the particle diameter, $\varepsilon$ the turbulent dissipation rate, $\mu$ the dynamic viscosity of the carrier phase, and $\rho_f$, and $\rho_p $ are the carrier and particle density, respectively. 

The snapshots files can be found in the database labeled as:\\ \texttt{RM-2008-LIGHT-512.St6.XX.h5}, which have particles with $\tau_p =0.048282$, and  $\beta=3\rho_f/(\rho_f+2\rho_p)=0$, i.e., $\rho_f\ll\rho_p$. More details of the numerical setup ($Re_\lambda \approx 185$) can be found in \cite{bec2010turbulent,bec2010intermittency}. The experimental and simulation parameters are summarized in tables \ref{tab:par}, and \ref{tab:drops}.

\begin{table}
	\begin{center}
		\centering

		\begin{tabular}{ c c c c c c c c c c}
			\toprule
			Dataset & Grid Mode & $Re_\lambda$ & $St_\eta$  & $\varepsilon \mathcal{L}^4 / \nu^3$ & $D_p/\eta$&$\mathcal{L}/\eta$ & $\lambda/\eta$ & $\phi_v$ & $\rho_p/\rho_f$\\
			\midrule
			EXP-2D-AG-A & Random &250 &  0.9 & 4.3 $\times 10^{8} $ & 0.125&   175 &  35& 1.2 $\times 10^{-5}$ &800\\
			EXP-2D-AG-B& Random  &250 & 0.9 & 4.3 $\times 10^{8} $ &  0.125&  175 &  35& 2.3 $\times 10^{-5}$ & 800 \\
		 EXP-2D-OG-A&Open  & 30 & 0.05 & 9.0 $\times 10^{4} $ &  0.032&  20&  7& 1.2 $\times 10^{-5}$ &  800 \\
 	     EXP-2D-OG-B&Open  & 30 & 0.05 & 9.0 $\times 10^{4} $ &  0.032&  20&  7& 2.3 $\times 10^{-5}$ &  800 \\
		DNS \cite{bec2010turbulent,bec2010intermittency}& - & 185 & 1 & 1.1$\times 10^{10}$ &- & 314 & 26 & - & $\gg 1$ \\
			\bottomrule	\end{tabular}
		\caption{Main turbulence parameters for experiments and DNS data used on this work. $\lambda=u^\prime\sqrt{15\nu/\varepsilon}$ is the Taylor length scale, and $\nu \sim 1.5\times 10^{-5}\, [m^2s^{-1}]$ is the air viscosity. $u^\prime$ is the rms of the streamwise velocity fluctuations. The turbulent Reynolds number is defined as $Re_\lambda=u^\prime\lambda/\nu$ and $D_p$ is the value of the most probable diameter which was used to compute the Stokes number $St_\eta=\frac{(D_p/\eta)^{2}}{36}(1+2\rho_p/\rho_f)$, see \cite{sumbekova2017preferential}. $\rho_p/\rho_f$ is the density ratio between the particles, and the carrier phase. $\varepsilon$, and $\mathcal{L}$ are the carrier dissipation, and integral length scale, respectively. AG/OG stands for the random or open mode of the active grid. $\phi_v$ is volume fraction for the experimental data.  As numerical data is a `1-way' coupling point particle simulation, we assume it is very diluted $\phi_v\ll10^{-6}$ \cite{Balachandar2010}.
		More details on how the experimental unladen flow parameters have been calculated can be found on \cite{mora2019energy}.     \label{tab:par}}
	\end{center}
\end{table}

\begin{table}
	\begin{center}
		\centering

		\begin{tabular}{ c c c c}
			\toprule
			Dataset & $N_{snap}$ & $\langle N_p/N_{snap} \rangle$&$  \sigma_{\langle N_p/N_{snap} \rangle}$  \\
			\midrule
			EXP-2D-AG-A & 4500& 1 $\times 10^{3}$& 280 \\
			EXP-2D-AG-B & 4500& 5 $\times 10^{3}$& 1200 \\
			EXP-2D-OG-A & 4500& 3 $\times 10^{3}$& 330\\
			EXP-2D-OG-B & 4500& 1 $\times 10^{4}$& 800\\
			DNS & 30& 4 $\times 10^{4}$&0\\
			RPP & 1000& $10^{3}$,$\,10^{4}$,$\,10^{5}\,$&0\\
			\bottomrule	\end{tabular}
		\caption{Datasets summary.  $N_{snap}$ is the number of snapshots, $N_{p}$ is the number of particles,  $\langle N_p/N_{snap} \rangle$ is the average number of particles per snapshot, and $  \sigma_{\langle N_p/N_{snap} \rangle}$ is the standard deviation of the mentioned average. \label{tab:drops}}
	\end{center}
\end{table}

\subsection{Vorono\"{i} Tesselations}
\label{sc:vorot}
Originally, Monchaux et al. \cite{Monchaux2010} proposed the use of Vorono\"{i} tessellations to quantify preferential concentration in turbulent flows. In 2D, the Vorono\"{i} tessellation yields a collection of areas $A$ from particles spatial positions \cite{Ferenc2007}. Statistics can be then computed for this collection, e.g., the average area $\langle A\rangle$, or the area standard deviation $\sigma_A$.  The Vorono\"{i} analysis is usually conducted by considering the area collection normalized by its mean $\mathcal{V}=A/\langle A\rangle$, giving $\langle \mathcal{V} \rangle =1$. 

Monchaux et al. \cite{Monchaux2010} suggested that  Vorono\"{i} area PDFs can be characterized by their standard deviation ($\sigma_\mathcal{V}$), a conclusion they substantiated by the log-normal behavior present in these PDFs, and the collapse seen under the log-normal transformation for different cases studied (with different values of $Re_\lambda$, or concentration $\phi_v$). These observations led them to conclude that $\sigma_\mathcal{V}$ estimates the `intensity' of clustering. 

Then, to quantify `clustering', $\sigma_\mathcal{V}$ is compared with the respective standard deviation coming from particles spatially distributed following a RPP distribution ($\sigma_{RPP}$), which by definition has no spatial correlations at any scale.  Thus, preferential concentration is present if $\sigma_\mathcal{V}>\sigma_{RPP}$. It is clear that under this criterion, the voids (or area outliers of the area collection) contribute the most to the numerical value of $\sigma_\mathcal{V}$ \cite{sumbekova2017preferential}. 

\begin{figure}
		\centering
		\includegraphics[trim=1cm 6cm 1cm 6.0cm, clip=true,scale=0.5]{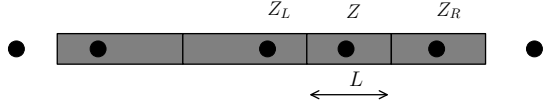}
		\caption{Undimensional Vorono\"{i} tessellation (1DVOA). For a given particle position $Z$ with left, and right neighbour particle $Z_L$, and $Z_R$
			respectively, the length of the Vorono\"i cell is given by $L=\vert Z_R-Z_L \vert /2$.}
		\label{fig:vsk}
	\end{figure}

Likewise, the unidimensional Vorono\"{i} tessellation (1DVOA) generates a collection of lengths $L$, not to be confused with the turbulence integral length scale $\mathcal{L}$ (figure \ref{fig:vsk}). However, Mora et al.\cite{Mora2018} reported that clustering could be present but not retrieved via 1D measurements \cite{shaw2003particle,kostinski2001scale}, and therefore, the criterion $\sigma_\mathcal{V}>\sigma_{RPP}$ cannot be directly extended to 1D measurements, and it may not be entirely conclusive for 1DVOA.

For clarity, we will define some acronyms below, but will often recur to the acronym's verbatim. First, angle brackets $\langle ... \rangle$ represent an ensemble average. Second: our short notation for the Vorono\"{i} tessellations variables goes as; $\mathcal{V}=\mathcal{M}_{voro}/\langle \mathcal{M}_{voro} \rangle$  where $\mathcal{M}_{voro}$ represents the length  $L$, area $A$, or volume $V$ of a Vorono\"{i} cell. Thirdly,
a 1D/2D/3D Vorono\"{i} tessellation analysis will be referred as 1DVOA, 2DVOA, or 3DVOA, respectively. Finally, the respective standard deviation of a RPP distribution ($\sigma_{RPP}$), which has no correlations at any scale, can be analytically computed: $\sigma_{RPP} \approx 0.71/ 0.53/ 0.42$ for 1DVOA/2DVOA/3DVOA, respectively (see table \ref{tab:fer}, and references~\cite{Ferenc2007,tanemura2003}). The 2D and 3D Vorono\"{i} diagrams were computed using the python library \textit{freud} \cite{freud}.

  \begin{table}[]
 	\begin{center}
 		\centering
 		\begin{tabular}{|c|c|c|c|c}
 			\cline{1-4}
 			Dimension & PDF expression & $\sigma_{RPP}$ & Model      &  \\ \cline{1-4}
 			1         &     $4\times\mathcal{V}^{1.0}e^{-2.0\mathcal{V}} $    & $\sqrt{1/2}$         & Analytical &  \\ \cline{1-4}
 			2         &      $343/15\sqrt{7/2\pi} \times\mathcal{V}^{2.5}e^{-3.5\mathcal{V}} $     & $\approx0.53$     & Fit        &  \\ \cline{1-4}
 			3         &   $345/7\times \mathcal{V}^{3.8}e^{-4.0\mathcal{V}^{1.17}}  $      & $\approx 0.42$     & Fit        &  \\ \cline{1-4}
 			\end{tabular}
 			\caption{ Summary of the Vorono\"{i} PDF expressions for 1,2,3 dimensions, and their respective properties as reported by Ferenc and Ned\'{a} \cite{Ferenc2007} (1D and 2D) and Tanemura \cite{tanemura2003} (3D).\label{tab:fer}}
 			\end{center}
 			\end{table}

\subsection{1D `Projected' Vorono\"{i} Analysis}
\label{sc:vdns}

Following the method proposed by Mora et al. \cite{Mora2018}, we projected particle positions coming from previous 2D and 3D data, which exhibit preferential concentration, orthogonally onto a line.

\textcolor{black}{
This approach aims to emulate the basic features of an Eulerian measurement: a fixed probe in space (e.g. downstream from the grid of a wind tunnel). Thus, the particle coordinates orthogonal to the bulk velocity ($U_\infty$) were `frozen’ as the particles were advected by the flow figure \ref{fig:proj2}).} 

For the experimental 2D data sets, this projection process will be denoted as $ 2D_{EXP} \rightarrow 1D_{\perp}$, where $2D_{EXP}$ stands for 2D experimental images, and $1D_{\perp}$ is a uni-dimensional orthogonal projection into 1D, i.e., onto the streamwise axis $\gamma$ (figure \ref{fig:proj2}). For the numerical 3D data, the notation is ($3D_{DNS}\rightarrow1D_{\perp}$), where $3D_{DNS}$ is the 3D dimensional DNS data set (figure \ref{fig:proj3}).  We applied this procedure for each data snapshot (table \ref{tab:drops}).

Thus, our `projection' algorithm \cite{Mora2018} comprised the following steps:

\begin{figure}
\centering
\begin{subfigure}{0.48\textwidth}
		\includegraphics[scale=0.9]{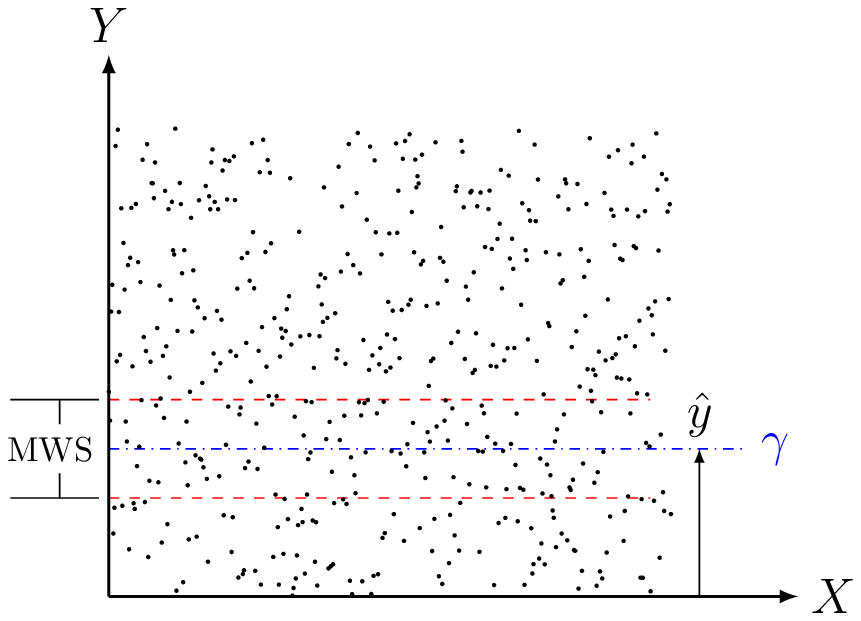}
		\caption{ \label{fig:proj2}}
\end{subfigure}
\quad
\begin{subfigure}{0.48\textwidth}
		\includegraphics[scale=0.6]{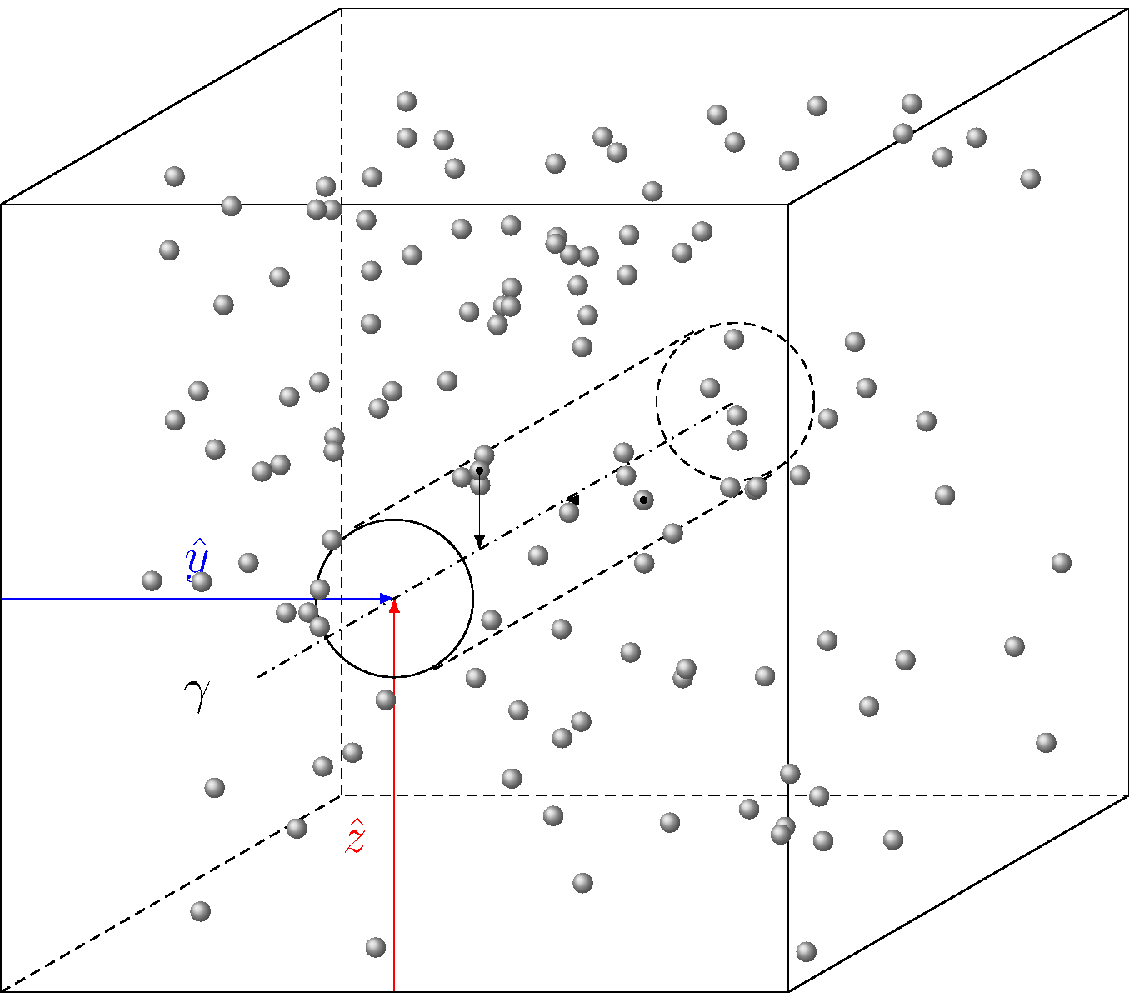}
		\caption{ \label{fig:proj3}}
\end{subfigure}
\centering
\begin{center}
\begin{subfigure}{\textwidth}	
	\includegraphics[scale=.45]{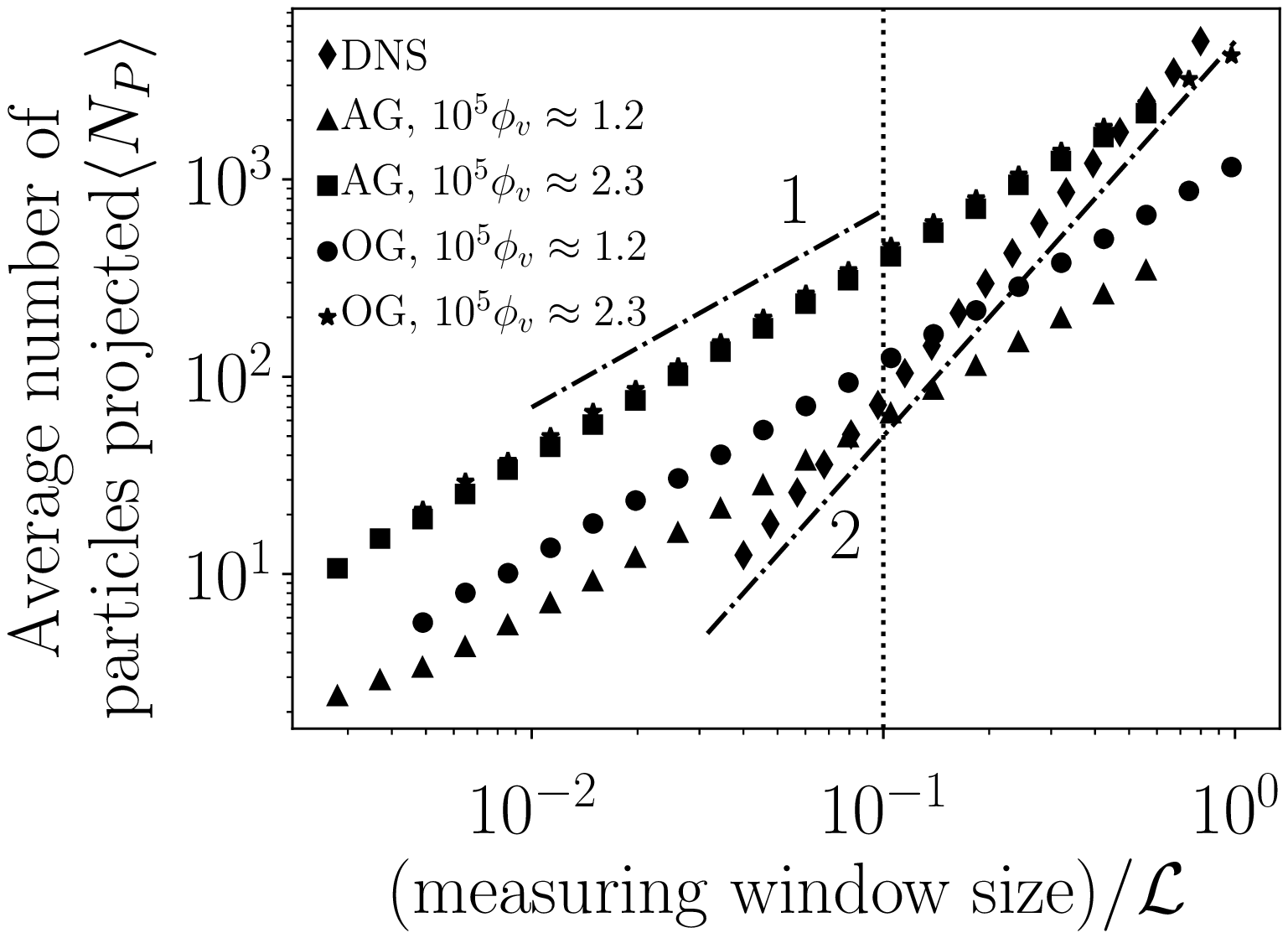}	
	\caption{ \label{fig-nss}}
\end{subfigure}
\end{center}

\caption{a) Sketch of the $2D_{EXP}\rightarrow1D_{\perp}$ particles centers projection for an arbitrary image. MWS is the measuring window size, $\hat{y}$ is the randomly generated  vertical coordinate of the axis $\gamma$ over which the points are orthogonally projected. b) Sketch of the $3D_{DNS}\rightarrow1D_{\perp}$ particles centers projection for an arbitrary DNS snapshot. MWS is the measuring window size equal to the cylinder diameter, $\hat{y}$ and $\hat{z}$ are the randomly generated coordinates of the axis $\gamma$ onto which the points are orthogonally projected. c) Average number of projected samples per snapshot from 1D sampling ($2D_{EXP}$ or $3D_{DNS} \rightarrow 1D_\perp$), and its dependency with the measuring window size. At very small MWS with respect to $\mathcal{L}$ the average number of samples captured is small, which is directly linked to lack of clustering recently reported \cite{Mora2018}.}
\end{figure}

\begin{enumerate}
	\item A random vertical $\hat{y}$ coordinate was generated. This coordinate set the position of the axis $\gamma$ onto which the particles were projected. 
	\item A symmetric measuring volume size (MWS) window was defined intended to quantify the effect of the instrument finite spatial resolution. At all times, the $\hat{y}$ coordinate choice was restricted so that the MWS width was inside the computational domain.
	\item All the particle \textit{centers} that lay within this measuring strip (axis $\gamma$ and width MWS) were projected onto the axis $\gamma$, i.e., their horizontal coordinate was recorded.		
	\item 1DVOA was performed over the particle location line projections.
	\item Three additional elements were considered for the 3D data : 
	\begin{enumerate}
	    \item A $\hat{z}$ transverse coordinate was generated to position a cylinder axis.
	    \item The generated cylinder diameter was equal to the measuring window.
	    \item All particles within this cylindrical volume were subsequently projected on the cylinder axis $\gamma$ (see figure \ref{fig:proj3}). 
	\end{enumerate}
	 
\end{enumerate}

The previous algorithm seems to capture the basic features of an equivalent experimental eulerian measurement, and is suitable to examine the effect of different measuring window sizes have on 1D Vorono\"{i} statistics. For instance, most quasi-unidimensional instruments \cite{hong2004characterization,cartellier1998monofiber,bachalo1984phase} are particle `counters', which usually yield a list of events with their respective eulerian `arrival' time. One would then expect that the larger the window size (the region in space where the instrument detects the particle transiting), the larger the number of droplets is `detected', a behavior this `projection' approach does indeed capture (see figure \ref{fig-nss}).

\section{Results}
\subsection{PDFs and Vorono\"{i} cell standard deviation}
\label{sc:prc}

We first checked if preferential concentration was present in the 2D and 3D data. Using the same 2D datasets, Mora et al.\cite{Mora2018} and Obligado et al.\cite{obligado2019study} have found strong evidence of preferential concentration by means of planar Vorono\"{i} tessellations. Next, we verified the existence of clustering in the 3D numerical database \cite{bec2010turbulent,bec2010intermittency}
by means of tridimensional Vorono\"{i} Tessellations~\cite{Ferenc2007}.

\begin{figure}
	\centering
\begin{subfigure}[h!]{0.48\textwidth}
		\includegraphics[scale=.4]{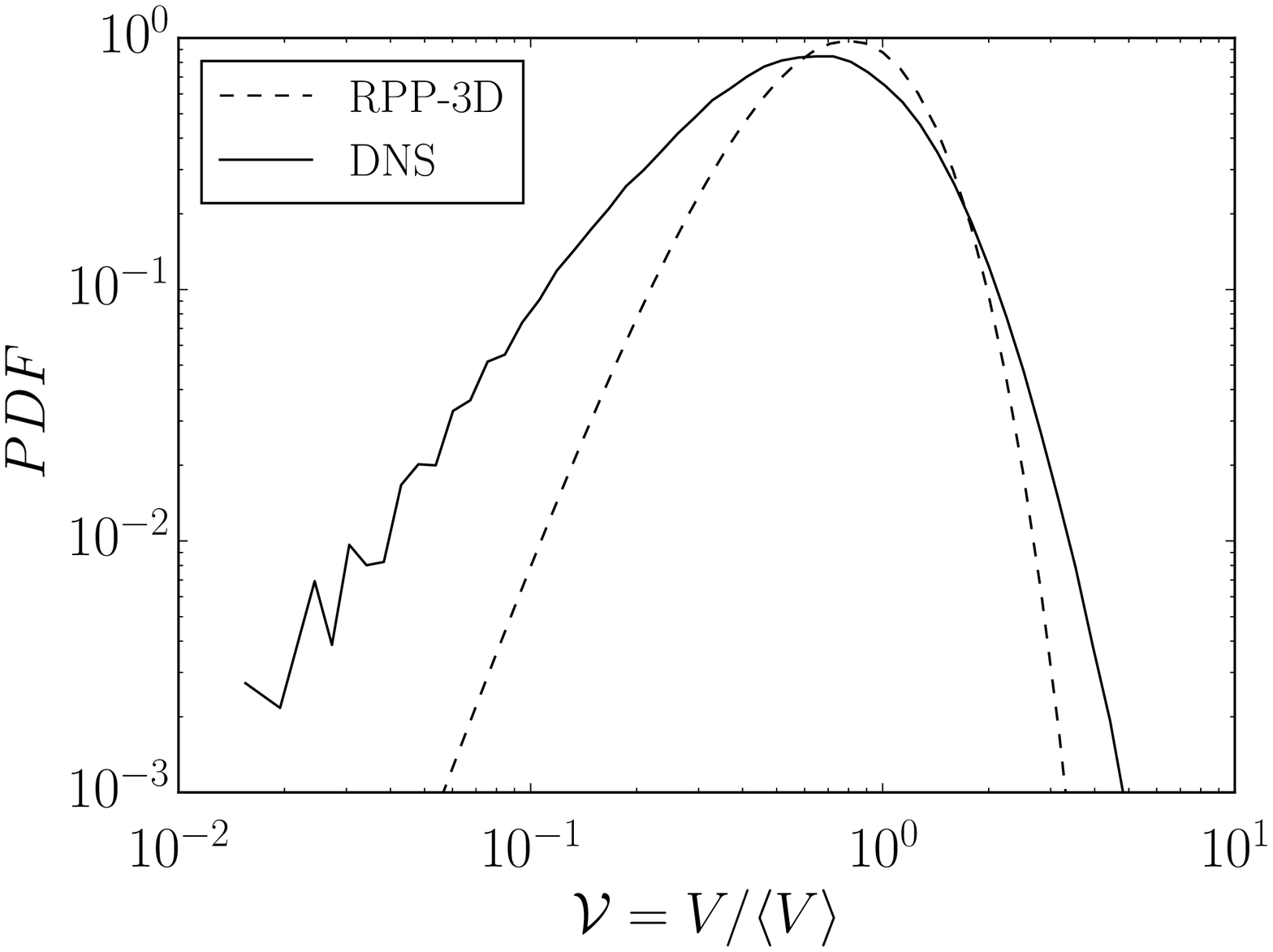}
		\caption{\label{fig-vors3D-DNS}}

\end{subfigure}
\quad
\begin{subfigure}[h!]{0.48\textwidth}
		\includegraphics[scale=.5]{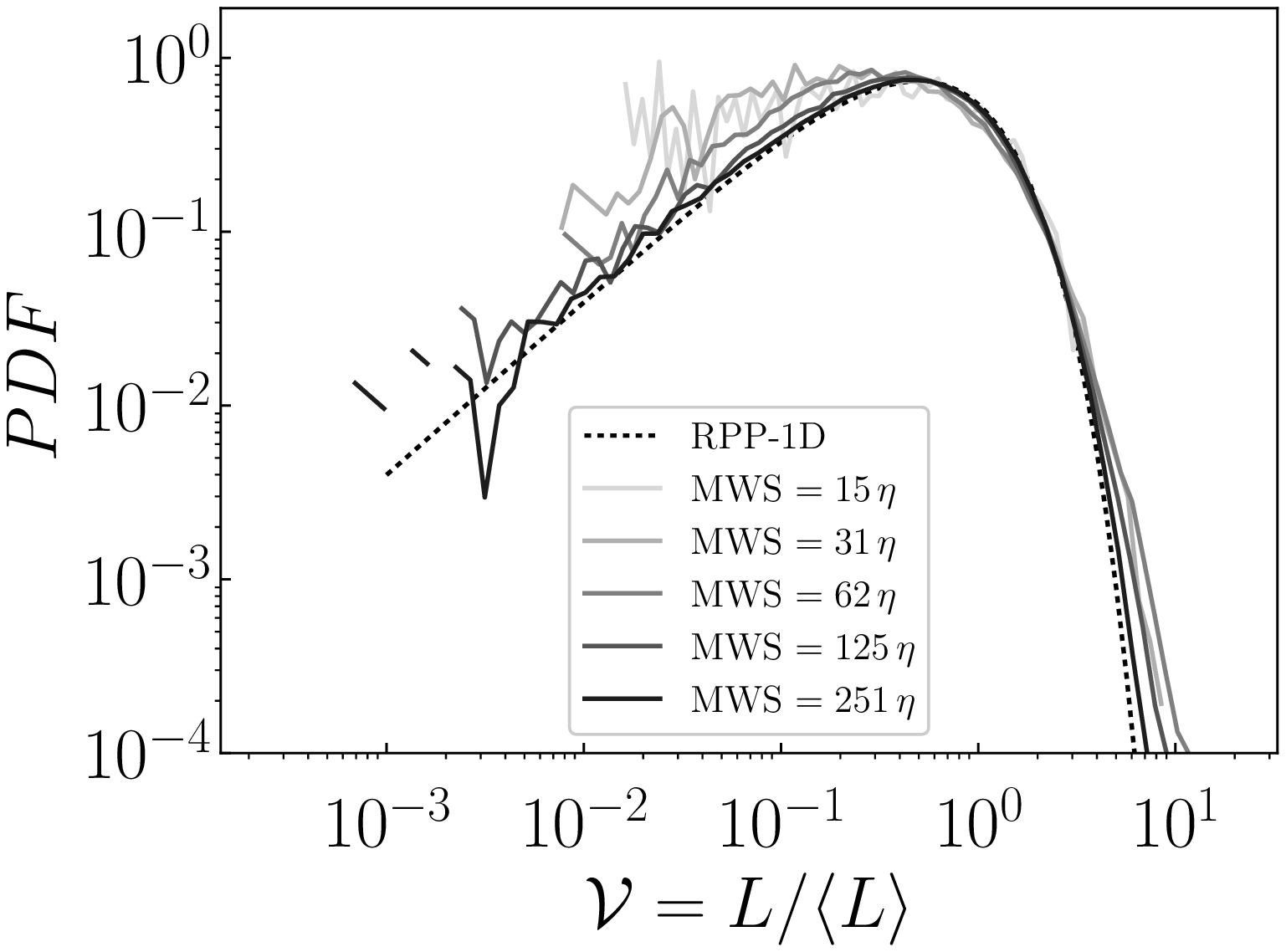}
		\caption{\label{fig-vors-DNS}}
\end{subfigure}
\caption{a) Probability density function plot of 3DVOA for the DNS data \cite{bec2010turbulent,bec2010intermittency}. Following the criterion of Monchaux et al. \cite{Monchaux2010}, it is clear that the DNS data contains clustering, as $\sigma_\mathcal{V} \approx 0.62>\sigma_{{3D_{RPP}}}\approx 0.42$ is larger than its equivalent one for a 3D RPP distribution. b) PDF plot of ($3D_{DNS} \rightarrow 1D_\perp$) 1DVOA for several MWS. }
\end{figure}

Visual inspection of the probability density function (PDF) (see figure \ref{fig-vors3D-DNS}) of the normalized Vorono\"{i} cell volume ($\mathcal{V}=V/\langle V\rangle$)  suggests the presence of preferential concentration within the DNS data. This is further confirmed by the larger value of the standard deviation of $\mathcal{V}$ (following the criterion proposed by Monchaux et al. \cite{Monchaux2010}) with respect to the RPP distribution, which has no correlation at any scale (see section \ref{sc:vorot}), i.e., $\sigma_\mathcal{V} \approx 0.62 > \sigma_{{3D_{RPP}}}\approx 0.42$, (see table \ref{tab:fer}).

Having confirmed the presence of preferential concentration within the 3D data, we applied the projection algorithm to it (see section \ref{sc:vdns}). Then, we conducted a unidimensional Vorono\"{i} tessellation analysis to these  $3D_{DNS}\rightarrow1D_{\perp}$ projections.

The probability density functions of the Vorono\"{i} cells sizes $\mathcal{V}=L/\langle L\rangle$ computed for several measuring window sizes (MWS), reveal a clear trend on these PDFs with varying window size (figure \ref{fig-vors-DNS}): the larger the window size was, the closer the PDF shape approached the respective RPP distribution.

The right tail of the Vorono\"{i} PDFs ($\mathcal{V}\gg1$) exhibits large changes with varying measuring window size. The importance of this cannot be overstated given
that the criterion proposed by Monchaux et al.\cite{Monchaux2010} to determine the existence of clustering relies on the numerical value of standard deviation $\sigma_{\mathcal{V}}$, which in turn heavily depends \cite{sumbekova2017preferential} on the cell values much larger than the mean $\langle L\rangle$, i.e., $\sigma_{\mathcal{V}}=\int_0^\infty \big(\mathcal{V}-\langle\mathcal{V}\rangle\big)^2PDF(\mathcal{V})d\mathcal{V}$.

\begin{figure}
	\centering
\begin{subfigure}[h!]{0.48\textwidth}
		\includegraphics[scale=.48]{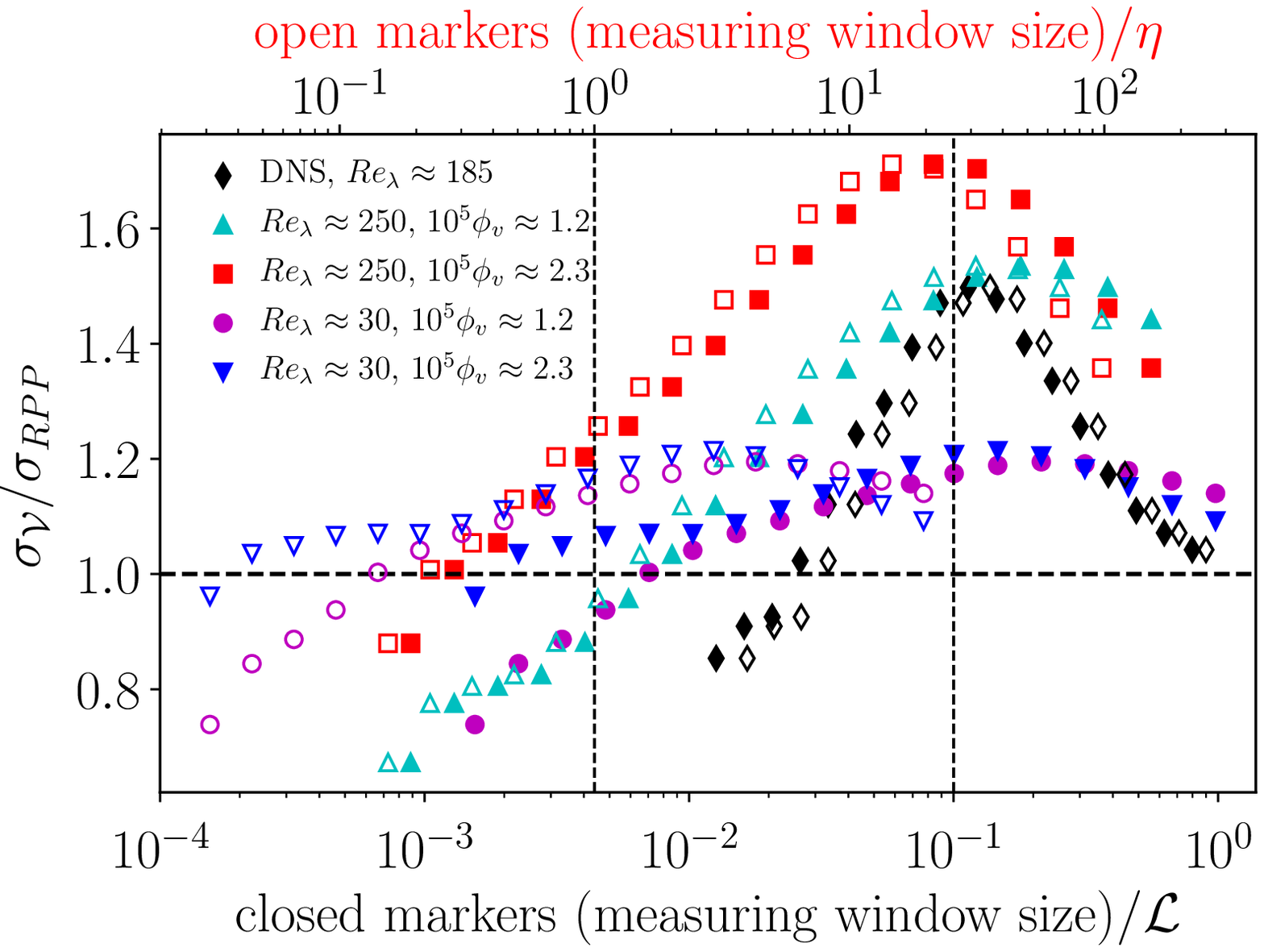}

		\caption{\label{fig-std-2dsc-L}}
\end{subfigure}
\begin{subfigure}[h!]{0.48\textwidth}

	\includegraphics[scale=.4]{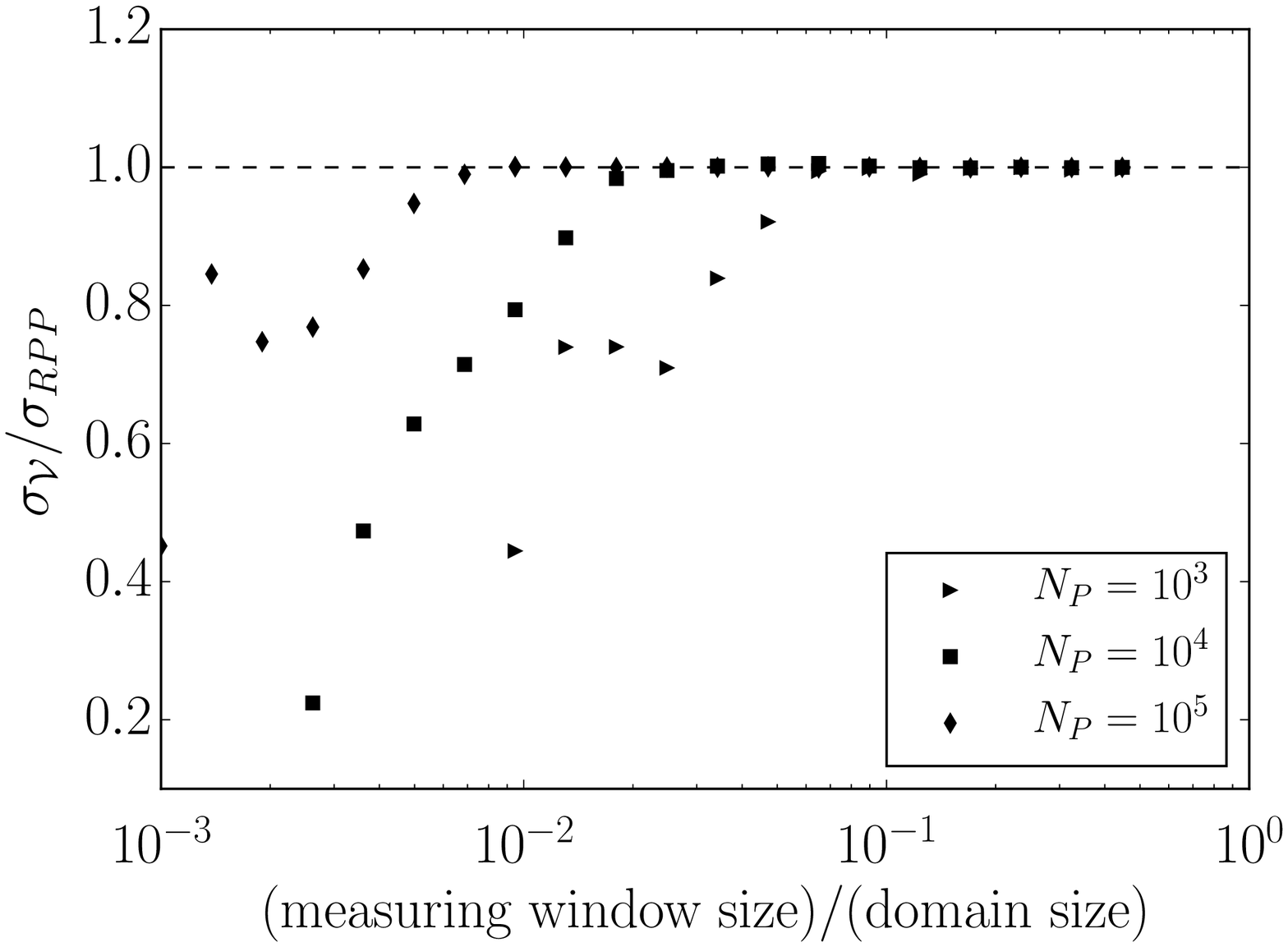}

	\caption{ \label{fig-std-3ds-mm}}
\end{subfigure}

\caption{ a) 1DVOA standard deviation evolution for the datasets used. The larger the concentration, the higher $\sigma_{\mathcal{V}}$ for fixed measuring window size. The peak location follows the relation MWS$_\star \approx \mathcal{L}/10$, where $\mathcal{L}$ the integral length scale of the flow (see table \ref{tab:par}). However, its value depends on $Re_\lambda$, as these and previous studies have shown \cite{sumbekova2017preferential} . b) 1DVOA standard deviation evolution of projections coming from synthetic random 3D data ($3D_{RPP} \rightarrow 1D_\perp$). $N_P$ stands for the number of points inside the 3D domain for 1000 synthetic snapshots.}

\end{figure}

Furthermore, when computing $\sigma_{\mathcal{V}}$ for our 2D, and 3D projections (figure \ref{fig-std-2dsc-L}), these results show that the presence of `clustering' via this criterion can only be recovered above a certain window size (MWS$_\dagger$), below which the evidence of clustering is not conclusive, i.e., $\sigma_{\mathcal{V}}/ \sigma_{{RPP}} \approx 1$. In fact, for some data, $\sigma_{\mathcal{V}}/ \sigma_{{RPP}}<1$ could lead to the wrong conclusion of a more ordered sub-poissonian distribution \cite{shaw2003particle,kostinski2001scale,Mora2018}. 

Given the smaller number of samples available with the projection method at smaller window sizes (see figure \ref{fig-nss}), one could attribute the result $ \sigma_{\mathcal{V}}/ \sigma_{{RPP}}<1$ to insufficient statistical convergence. On the contrary, having statistical convergence does not necessarily guarantee that the spatial correlations within the data are captured \cite{kostinski2001scale}.  

\begin{figure}
\begin{subfigure}[h!]{0.48\textwidth}
		\includegraphics[scale=.4]{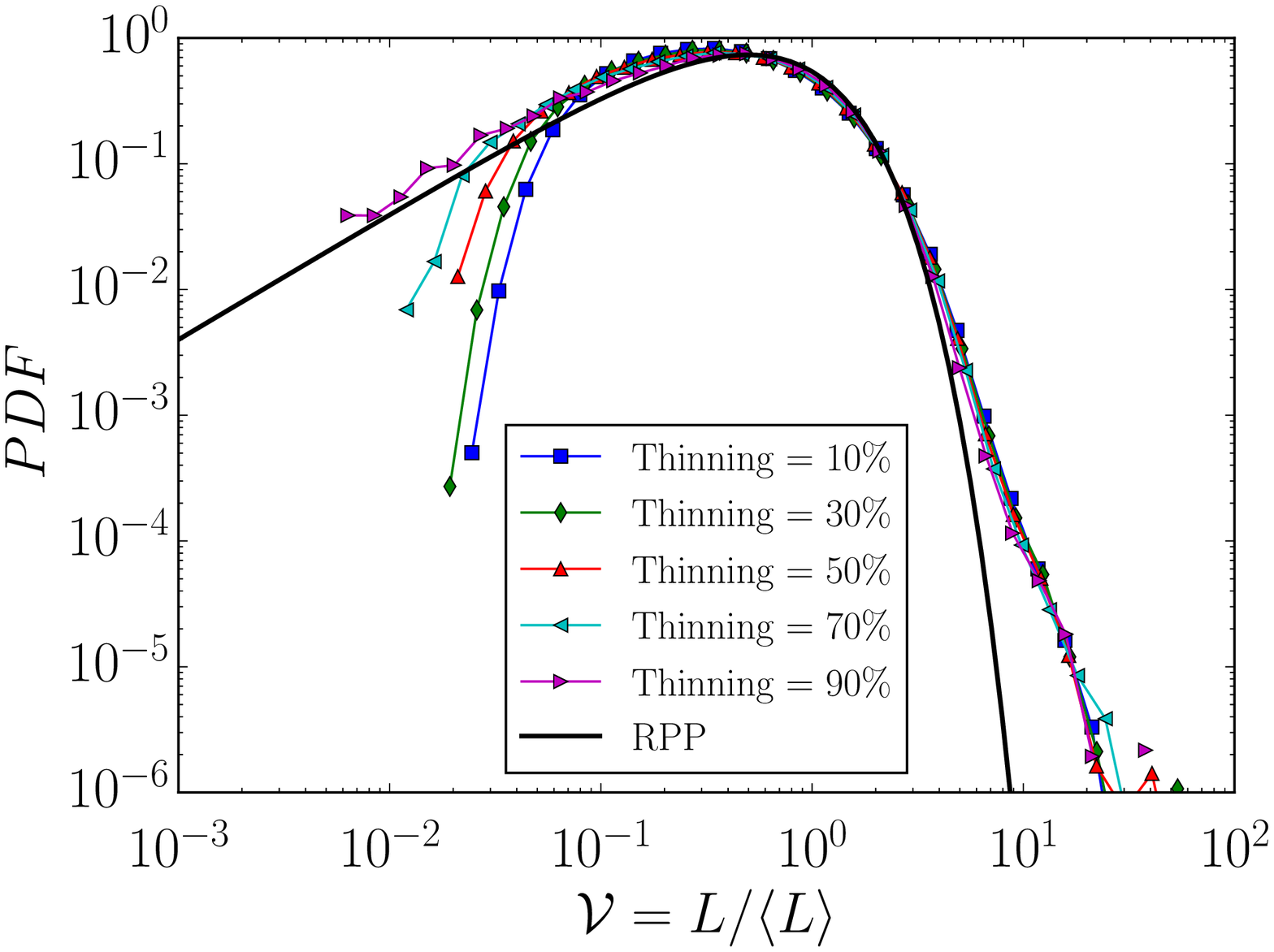}
		\caption{\label{fig-2d-rth}}
\end{subfigure}
\begin{subfigure}[h!]{0.48\textwidth}
	\includegraphics[scale=.4]{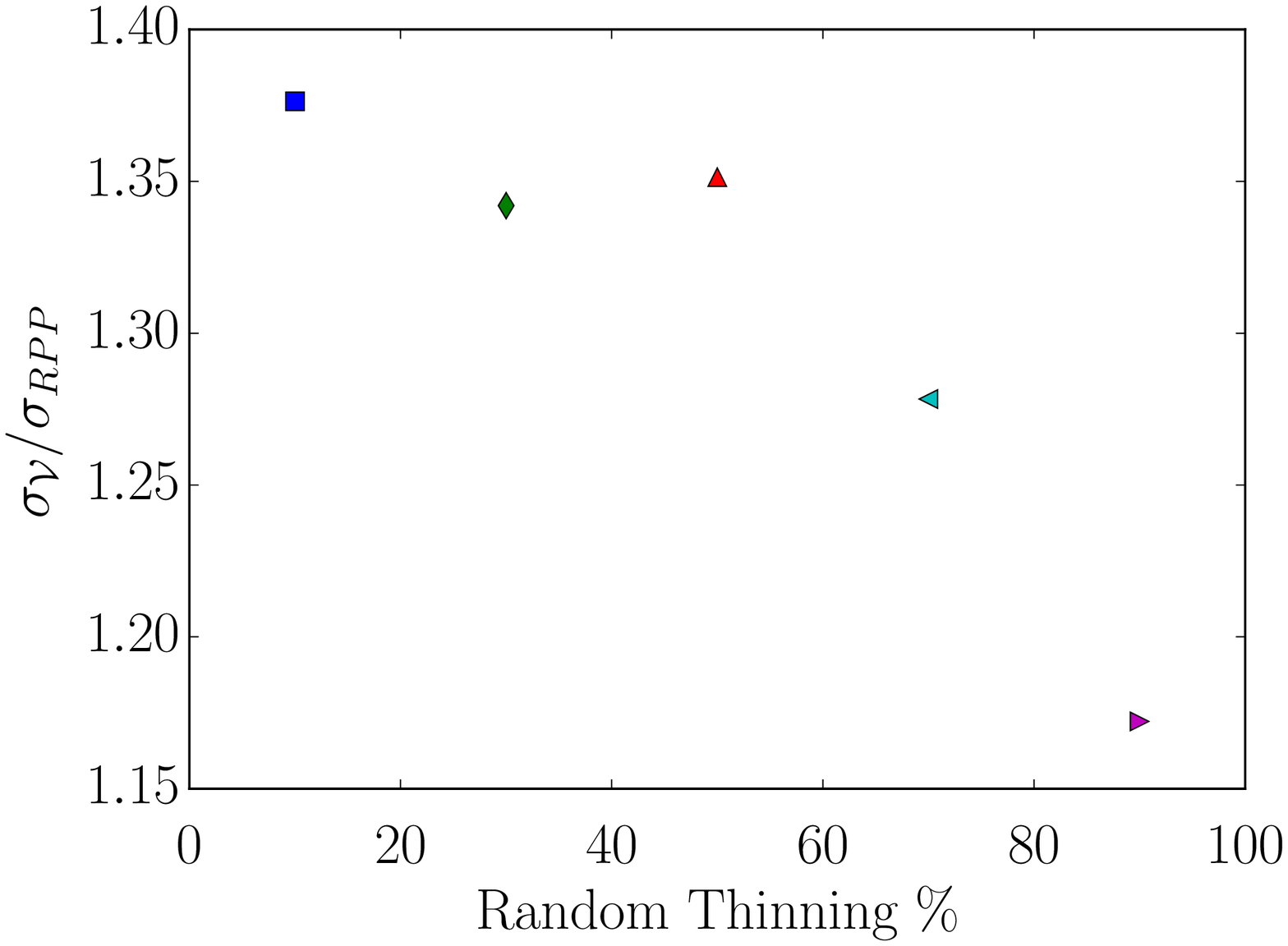}
		\caption{\label{fig-2d-rth-std}}
\end{subfigure}

\caption{ 1D PDI data from Sumbekova \cite{Sumbekova2016} for varying degrees of thinning (randomly removing samples from their records). The larger the percentage the larger the number of original particles removed. a) Thinned Vorono\"{i} PDFs b) Their corresponding normalised standard deviation ($\sigma_\mathcal{V}/\sigma_{RPP}$).}
\end{figure}

The latter can be illustrated by applying the same projection algorithm to a three-dimensional random set (labeled $3D_{RPP} \rightarrow 1D_\perp$ in table \ref{tab:drops}). Although this data set is, by definition, random, it shows a similar a transition region where $\sigma_\mathcal{V}/\sigma_{RPP}<1$ (figure \ref{fig-std-3ds-mm}). These data sets attain their `theoretical' value of $\sigma_\mathcal{V}$ after a certain size which depends on to the number of particles in the domain (concentration). Hence, it is the instrument and its capacity to capture particle spatial correlations which ultimately determines to which extent the 1DVOA is successful. Accurately identifying these correlations yields the proposed picture of clusters and voids \cite{Monchaux2010}, as there cannot be voids without clusters. Moreover, if evidence of clustering ($\sigma_\mathcal{V}/\sigma_{RPP}>1$) is retrieved by means of 1DVOA, it is reliable, as a strong signature of clustering is present even after randomly removing 70$\%$ of the data points (figures \ref{fig-2d-rth} and \ref{fig-2d-rth-std}).


The previous observations are very relevant from an experimental point of view, as they aid to explain why two distinct measuring instruments with dissimilar window sizes could yield opposite results; as seen in the experiments of Sumbekova \cite{Sumbekova2016} and Mora et al. \cite{Mora2018}, ran under almost identical conditions but with different 1D measuring instruments. For the former experiment, a PDI device (MWS $\approx \eta$) was used, while for the latter, optical probes were used (MWS $ \approx 0.1\eta$).
This order of magnitude difference in the instrument window size is key to understand the origin of this contradiction (figure \ref{fig-std-2dsc-L}).

These results, therefore, indicate that a sensitivity analysis has to be performed if a unidimensional Vorono\"{i} analysis retrieves no clustering where higher dimensional techniques do (2DVOA or 3DVOA). However, considering the observed strong dependency of the minimum window size (MWS$_\dagger$) on the concentration values $\phi_v$ for which $\sigma_\mathcal{V}/\sigma_{RPP}>1$ (figures \ref{fig-std-2dsc-L} and \ref{fig-std-3ds-mm}), and the strong dependency of $\sigma_\mathcal{V}/\sigma_{RPP}$ on the carrier phase conditions $Re_\lambda$, such sensitivity analysis is not straightforward, and it should be tuned to the specific conditions of the experiment. 

Interestingly with our projection algorithm, we also recovered that the degree of clustering increases with $Re_\lambda$ in agreement with previous studies \cite{Obligado2014,sumbekova2017preferential}. 
\textcolor{black}{Our results, and the data of Sumbekova \cite{Sumbekova2016}  suggest that an instrument with a measuring volume equal or larger than the Kolmogorov length scale ($\eta$) should be used if evidence of preferential concentration is to be retrieved by means of 1D Vorono\"{i} tessellation analysis. Likewise by means of RDF functions, Saw et al. \cite{Saw2012} have recovered evidence of preferential concentration using a probe with a volume of order $\eta$.}

Although the choice of $\eta$ as MWS$_\dagger$ could be seen as arbitrary, it seems to be adequate for the current range of $Re_\lambda$ attainable in several experimental facilities (and in our data sets) under similar experimental conditions. More importantly, this criterion does not contradict previously reported clustering at scales smaller than $\eta$ \cite{Toschi2009}. If these correlations were to be present at smaller scales than $\eta$, they would be effectively captured (figure \ref{fig-std-2dsc-L}). 

Despite being only intended to capture the basics of an eulerian measurement, our approach also seems to capture interesting information related to preferential concentration physics, giving extra credence to its underlying hypotheses.

On the one hand, it is remarkable that the peak of the ratio $\sigma_\mathcal{V}/\sigma_{RPP}$ (maximum clustering degree in figure \ref{fig-std-2dsc-L}) occurs at a measuring window size close to a tenth of the integral length scale, i.e., MWS$_\star \approx \mathcal{L}/10$. This peak is representative of the maximum spatial correlation present at the data set, as our projection procedure does not increase the actual correlation in the data (see 3D RPP projections results in figure \ref{fig-std-3ds-mm}). Hence, this window size, or scale, can be connected to the multi-scale mechanisms proposed to explain preferential concentration \cite{Goto2006,ireland2016effect}.

For instance, the sweep-stick mechanism \cite{Goto2006,Goto2008,Coleman2009} considers that the scaling behavior for voids (the counterpart of clusters) in the inertial range is not described by a single scale, but instead follows a self-similar behavior within a range of scales (window size), $\ell_{max}/\ell_{min}=\mathcal{O}(10)$ (where $\ell$ denotes scales from the inertial range of the turbulent flow). We can therefore interpret  MWS$_\star$ as representative of the maximum interaction among all scales between the carrier turbulent flow and the particles. Also, the study of Bragg et al. \cite{bragg2015mechanisms} reports that their proposed mechanism responsible of clustering attains its peak at a scale close to $r=200\eta$, and with $\mathcal{L}=800\eta$, these observations are in the same order of magnitude of MWS$_\star \approx \mathcal{L}/10$.

\subsection{Average clusters size and measuring window size}
\label{sc-cs}

\begin{figure}
\centering
\begin{subfigure}[h!]{0.48\textwidth}
			\includegraphics[scale=.46]{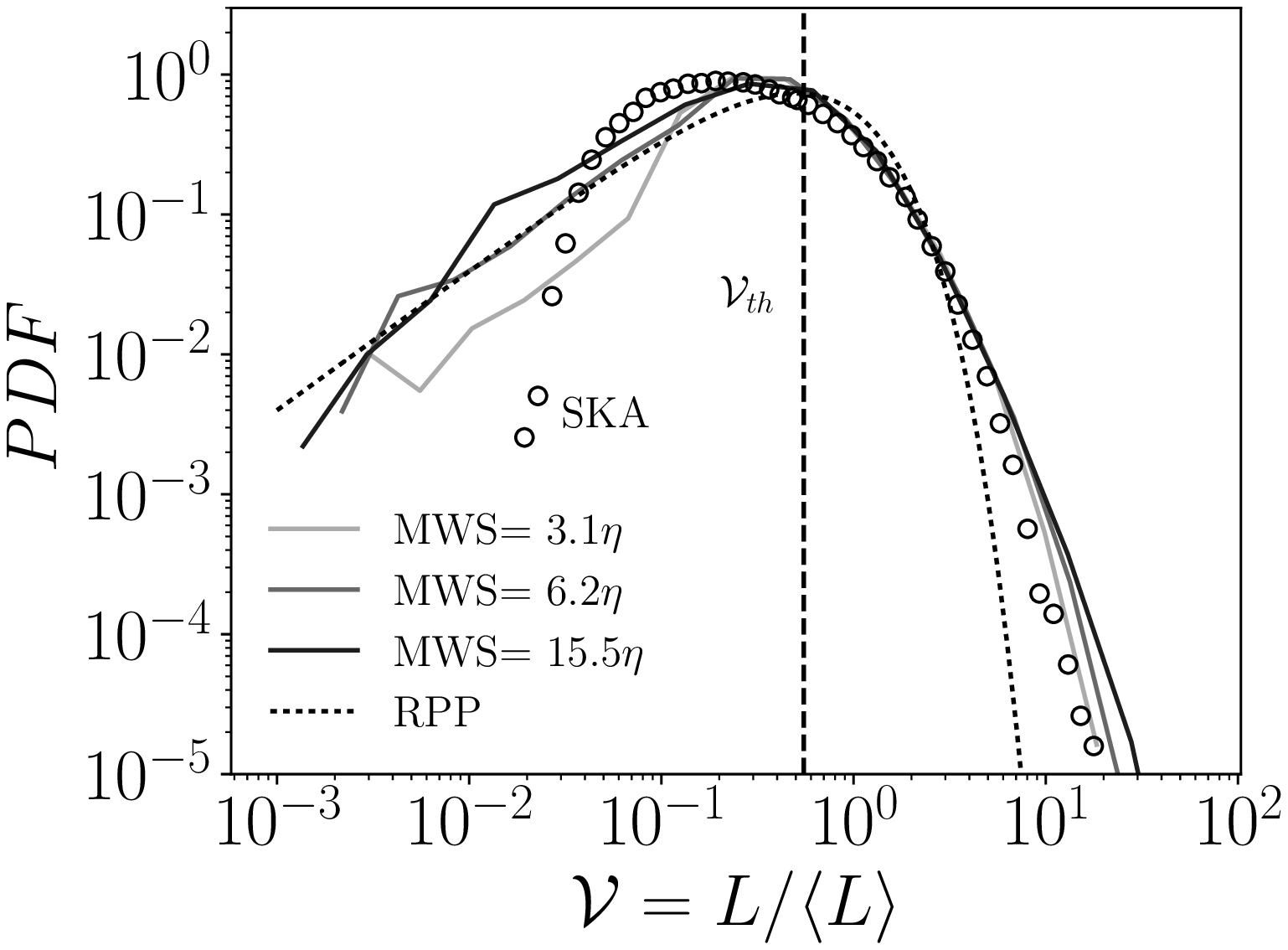}
			\caption{\label{fig-vors}}
\end{subfigure}	
\quad
\begin{subfigure}[h!]{0.48\textwidth}
		\includegraphics[scale=.4]{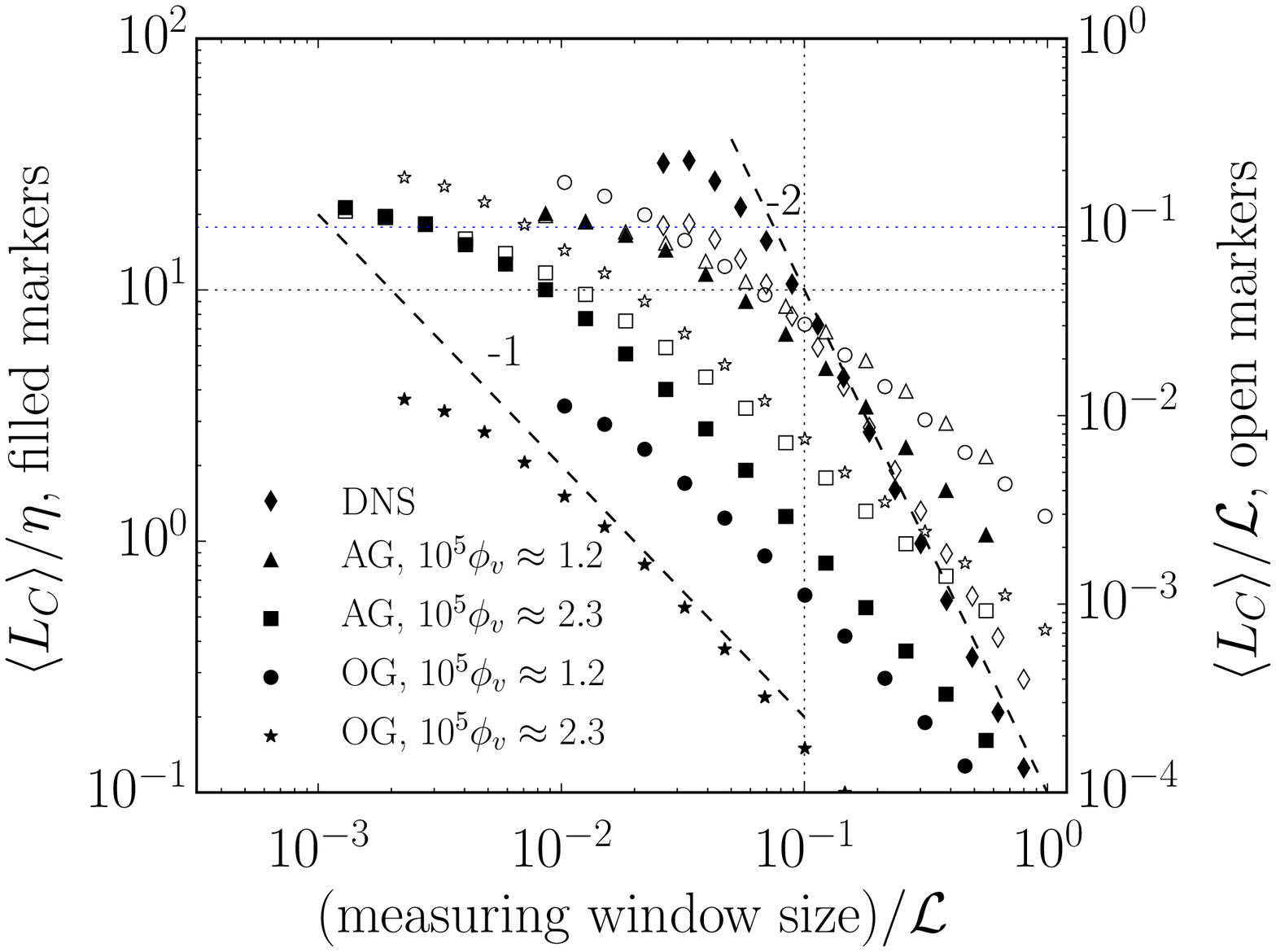}
		\caption{ \label{fig-cll}}
\end{subfigure}	
\caption{a) Probability density functions (PDFs) of ($2D_{EXP} \rightarrow 1D_\perp$) 1DVOA (EXP-2D-AG-B) for three different measuring window sizes (MWS). The $\circ$ (SKA) markers correspond to the data of Sumbekova \cite{Sumbekova2016} for $10^{5}\phi_v=2$. b)  Average linear cluster size $\langle L_C\rangle$ vs different measuring window sizes (MWS) for the data analyzed. The filled markers represent the axis on the left where the cluster average size is normalized by $\eta$. The open symbols correspond to the axis on the right, where the cluster average size is normalized by $\mathcal{L}$ (the integral length scale).}
\end{figure}

We will now examine the effects that the observed biases have on the cluster linear size $L_C$. We started by computing $L_C$ following the most widely accepted cluster identification algorithm \cite{Monchaux2010}: first we selected all the normalized Vorono\"{i} cell sizes $\mathcal{V}$ that were below a threshold value, $\mathcal{V}<\mathcal{V}_{th}$. This threshold is defined as the closest crossing (to the left of the RPP peak) between the RPP PDF, and our data PDF (see figure \ref{fig-vors}), i.e., $\mathcal{V}_{th}=\mathcal{V}=\mathcal{V}\vert_{RPP}<1$ \cite{Monchaux2010,Monchaux2012,Sumbekova2016a}. 

The second step in the cluster algorithm is to find (within the collection $\mathcal{V}<\mathcal{V}_{th}$) cells which share at least one edge. Then, two or more neighbouring cells were considered to be a cluster ($N_{PC}\geq 2$, where $N_{PC}$ is the number of particles inside the cluster).

We took the same Vorono\"{i} cell threshold ($\mathcal{V}_{th}\approx 0.55$) for all 2D and 3D projections. It is important to note that this cluster identification algorithm is applied individually to each snapshot (see section \ref{sc:vdns} and table \ref{tab:drops}). Otherwise, spurious results in the different metrics, not shown here, could arise. 

The results (see figure \ref{fig-vors} \& \ref{fig-cll}) for the different data sets, and constrained to the MWSs for which $\sigma_{\mathcal{V}}/\sigma_{RPP}>1$, show two distinct regions: one where the average cluster size has a power-law decay behavior with increasing measuring window, and another where $\langle L_C \rangle /\eta$ evolves slowly with MWS.

The former region occurs at larger window sizes than the minimum for which $\sigma_\mathcal{V}/\sigma_{RPP}>1$ (see in figure \ref{fig-std-2dsc-L}, the region where MWS$\gg$MWS$_\dagger$, the minimum window size). The power law behavior is a direct result of the projection method, and can be easily explained if one takes into account that: first, the average Vorono\"{i} cell value $\langle L\rangle$, which is inversely proportional to the particle number ($N_p$), i.e., $\langle L\rangle\propto N^{-1}_p$, also exhibits an analogous behavior with the window size (see figure \ref{fig-nss}). Second, it has to be noted that the Vorono\"{i} cells below the threshold ($\mathcal{V}<\mathcal{V}_{th}$) are the ones that contribute the most to $\langle L\rangle$ \cite{sumbekova2017preferential}.  Hence, it is not surprising that clusters become smaller (and denser) as the concentration is locally increased by projecting more particles onto a fixed domain; a leap in concentration yields a proportional reduction in $\langle L_C \rangle$, a potential setback on the applicability of 1DVOA for cluster characterization.

Given these considerations, we proceed to analyze the transition region for which $\langle L_C \rangle$ varies slowly with MWSs. The extent of this region seems heavily dependent on the liquid fraction $\phi_v$ and $Re_\lambda$. More importantly, the estimated average cluster lengths ($\langle L_C\rangle/\eta$) are again in agreement with previous 2D and 3D studies. \textcolor{black}{Although early research shows \cite{Aliseda2002,Obligado2014,Bateson2012,uhlmann2017clustering,monchaux2017settling} that average cluster size is of order 10$\eta$ for $Re_\lambda=O(100)$, recent research has shown \cite{wittemeier2018explanation,sumbekova2017preferential,petersen2019experimental} that $L_C$ may grow for larger values of $Re_\lambda$, for instance, for $Re_\lambda=O(300)$ \cite{Obligado2014,petersen2019experimental,sumbekova2017preferential} report $L_C=O(20-100\eta)$ in agreement with our observations. }

In fact, the data (figures \ref{fig-std-2dsc-L} and \ref{fig-cll}) suggest that the maximum clustering window size MWS$_{\star} \approx \langle L_C\rangle \approx  0.1\mathcal{L}$. This conjecture seems to be supported by previously published data, for instance; Monchaux and Dejoan \cite{monchaux2017settling} reported $\langle L_C\rangle/\eta\ \approx 2-4$, with $\mathcal{L}/\eta \approx 30$, and $Re_\lambda \approx30$, which is in the same order of magnitude of $\langle L_C\rangle/\eta \approx 0.1\mathcal{L}/\eta$; Obligado et al. \cite{Obligado2014} and Sumbekova et al. \cite{sumbekova2017preferential}  outlined $\langle L_C\rangle/\eta\ \approx 10-20$ with $\mathcal{L}/\eta \approx 500$ and $Re_\lambda\in[200,400]$; and recent experiments from Petersen et al. \cite{petersen2019experimental} reported $\langle L_C\rangle/\eta \approx 10-40$ with $\mathcal{L}/\eta \approx 500$ and $Re_\lambda \in[200,500]$, which are again in rough agreement with MWS$_{\star} \approx \mathcal{L}/10$. The 3D DNS projected data also follows a similar trend, i.e.,  MWS$_{\star} \approx \langle L_C\rangle \approx40 \eta$, which is close to the value found from 3D Vorono\"{i} tessellation analysis. 

Thus, the expression $\langle L_C\rangle\approx \mathcal{L}/10$ could constitute a scaling for the cluster characteristic size, under similar experimental conditions. This latter consequence is remarkable, and consistent with the view of the strong role of turbulence in controlling the spatial correlation in the particle concentration field \cite{Goto2008,Coleman2009}. For instance, despite having on average the same number of projected particles (see figures \ref{fig-nss} and  \ref{fig-cll}), the two 2D data sets with the highest concentrations, $\phi_v=2.3\times10^{-5}$, have characteristic sizes differing by almost an order of magnitude in agreement with previous 2D studies \cite{Obligado2014,sumbekova2017preferential}.

Once again, these results give credence to our conservative recommendation for a minimum window size close to $\eta$, as such probe will not only be able to recover evidence of preferential concentration, but also will retrieve similar values and trends of $\langle L_C\rangle/\eta$, under similar experimental conditions of $Re_\lambda$, and $\phi_v$.

\section{Cluster size PDF}
\label{sc-csc}

Having checked the effects of the measuring window size on the average cluster size $\langle L_C\rangle$, we proceed to study the sensitivity of the PDF of clusters length ($L_C$) to varying MWSs.

\begin{figure}[h!]
	\begin{subfigure}[h!]{0.48\textwidth}
		\includegraphics[scale=.5]{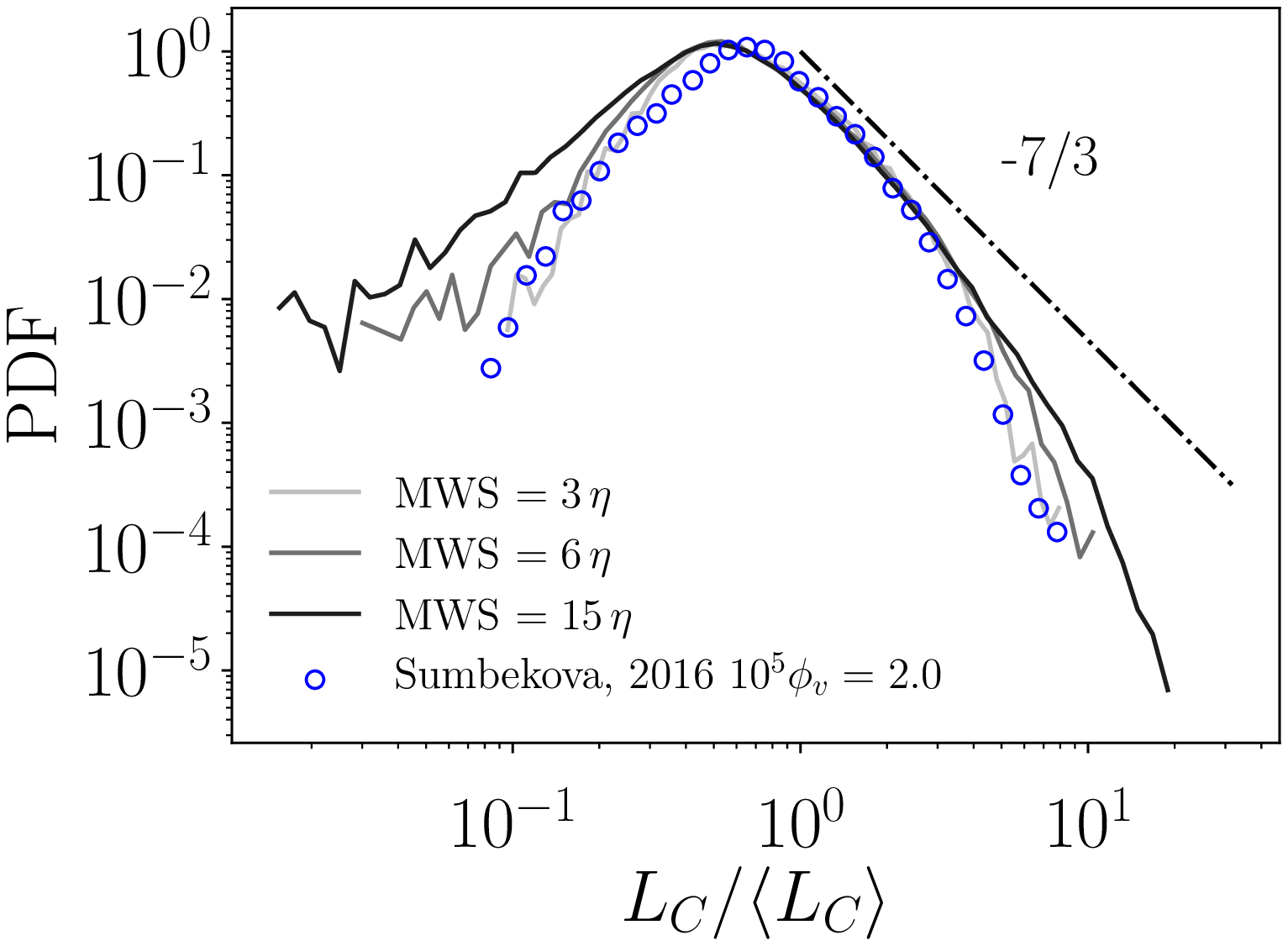}
		\caption{\label{fig-clu-l-ag}}
	\end{subfigure}
	\quad
	\begin{subfigure}[h!]{0.48\textwidth}
		\includegraphics[scale=.4]{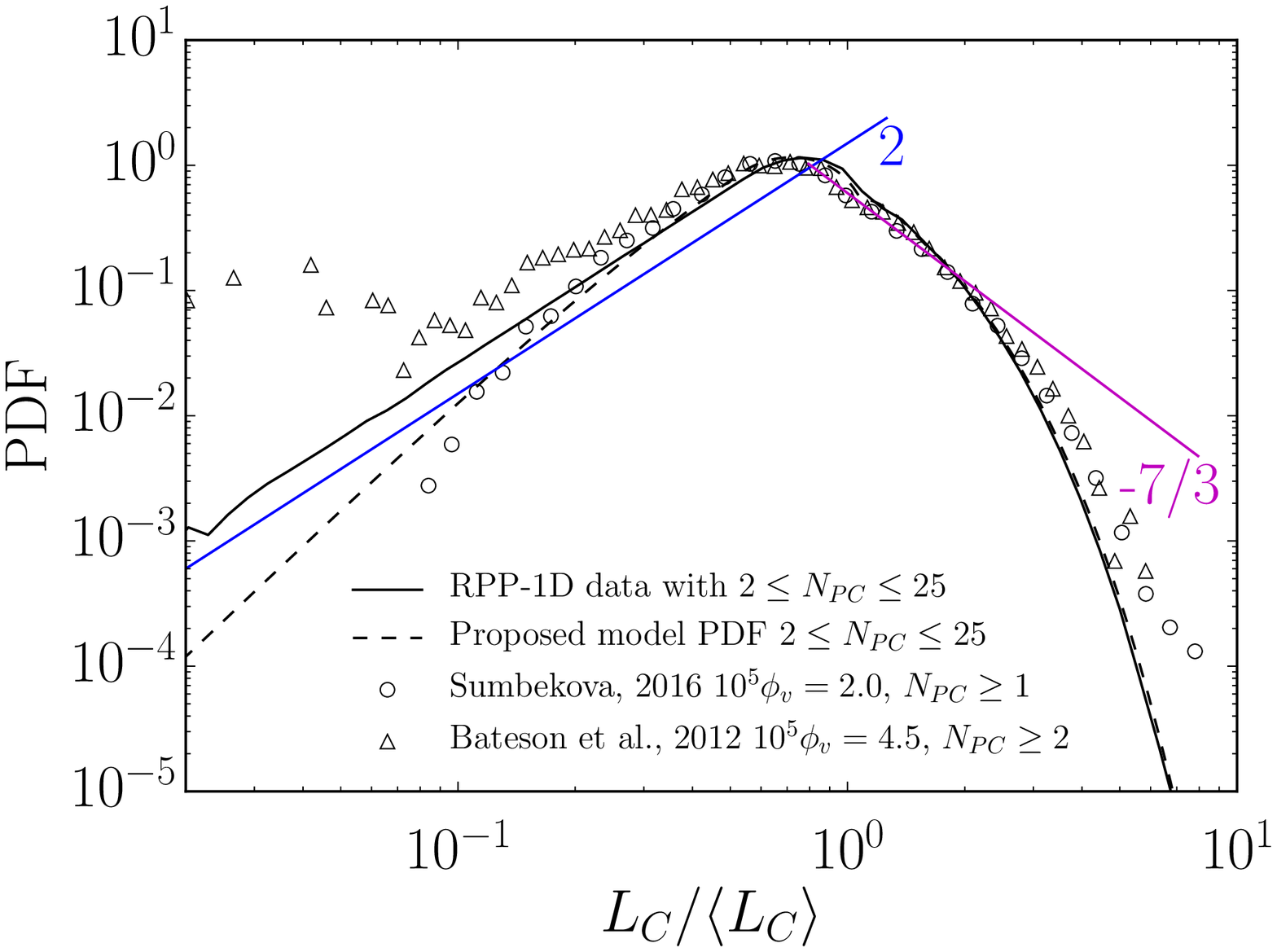}
		\caption{\label{fig-clu-rdn}}
	\end{subfigure}

	\caption{In the figures, the markers ($\circ$) and ($\triangle$) correspond to the experimental data from Sumbekova \cite{Sumbekova2016a}, and Bateson et al.\cite{Bateson2012} (taken in a different facility), respectively. a) PDFs of 1DVOA of linear cluster size  $L_{C}/\langle L_{C} \rangle$ for several window sizes (MWSs). b) PDFs of 1DVOA of linear cluster size  $L_{C}/\langle L_{C} \rangle$ for a random uniform distribution (RPP \cite{Ferenc2007}), experimental data, and the model proposed. For all clusters having between  2 and 25 cluster cells ($N_{PC}$).  The model  proposed here represents well the right tail of the RPP data when $N_{PC}\geq 2$. The condition $\mathcal{V}\leq\mathcal{V}_{th}=0.55$ was employed for clustering computation. }
	
\end{figure}

The analyses for experimental active grid data ($Re_\lambda\approx 250$, $10^5\phi_v\approx2.3$, the data from the other experimental conditions exhibited a similar behavior) yield two outstanding observations (see figures \ref{fig-clu-l-ag} and \ref{fig-clu-rdn}). First, the lack of a conclusive power law on the right tail of the $L_C$ PDF. Second, there is a rough collapse for the PDFs of all window size with previous experimental data.

The absence of a power law behavior is surprising, as it was previously found in multiple 2D and 3D Vorono\"{i} studies \cite{Obligado2014,monchaux2017settling,baker2017coherent,uhlmann2017clustering,huck2018role,petersen2019experimental}, and has been thought to be representative of the `turbulence-driven' clusters, as the arguments of Goto and Vassilicos \cite{Goto2006} suggest that the PDF of the voids areas should scale as $f_{AV}\sim A^{-5/3}$ based on Kolmogorov scalings, and a `-5/3' powerlaw in the velocity power spectral density. Remarkably, such power law is also found for the PDF of clusters area $A_C$, or cluster volume $V_C$ in both 2D and 3D. For our unidimensional records, such power-law behaviour barely exists (if at all), as it can only be identified for less than a decade for different datasets (figure \ref{fig-clu-rdn}) .

The second observation is somehow baffling, as one could not unambiguously anticipate that these projected data PDFs (numerical and experimental) would collapse at different MWSs. The existence of relative good agreement between the projected data and previous experiments is also intriguing \cite{Bateson2012,Sumbekova2016} (figure \ref{fig-clu-rdn}).

This observation has also been reported for the PDF of cluster volumes ($V_C$) in numerical simulations. For instance, Uhlmann and collaborators \cite{uhlmann2014sedimentation,uhlmann2017clustering,chouippe2019influence} have also found by means of 3D Vorono\"{i} tessellations that if the same clustering algorithm  (and with the same threshold values $\mathcal{V}_{th}$, see section \ref{sc:vorot}) is applied to data coming from a 3D random distribution, and to clustering containing data, both volume cluster PDFs attain (to some extent) a close resemblance. This pitfall is not minor, as it is critical to distinguish between the clustering coming from the turbulence and from the random fluctuations in the data. Otherwise, it is very difficult to correctly assess the influence of the turbulence on particle laden related variables such as particle settling velocity \cite{huck2018role}. 

Hence, we conducted a similar analysis as the one of Uhlmann's group to explore the origin of both observations: the lack of power law behavior, and the close PDFs resemblance in 1D.  Our approach was to apply the 1DVOA to synthetic data from a uniform random distribution with a uniform distribution of probability, and applying the same threshold value to those Vorono\"{i} cells with sizes smaller than 0.55, i.e., $\mathcal{V}<\mathcal{V}_{th}=$0.55. These cells were then processed using the cluster identification algorithm described in section \ref{sc-cs}. As random datasets have no actual clusters, the algorithm then detects the high concentration regions present on the RPP. This analysis will then be useful to compare these random fluctuations with the clusters generated by the interaction between inertial particles and a background turbulent flow.

The results of this analysis evince that the PDF of cluster sizes from 1D synthetic random data collapse remarkably well with the clusters PDFs generated from previous 1D experimental PDI particle records \cite{Sumbekova2016,Bateson2012} under turbulent conditions (see figure \ref{fig-clu-rdn}). As this covers most of the available clusters data, our next objective is to check whether these observations are a result of the Vorono\"{i} tessellation algorithm in 1D, or if they are a consequence of the cluster detection method \cite{Monchaux2010} in any dimension.

\subsection{Clusters PDF Model}
\label{sc:cpdf}

We developed an analytical model for cluster size probability distribution $f(L_C)$ coming from synthetic 1D random data (a detailed explanation can be found in the \ref{sc:Ap}). Our aim with this model was to gain insight into: a very weak (if existent at all) power law, and a rather universal shape for the cluster size PDF for any input 1D data fed into the 1DVOA.

We used a mixture PDF model \cite{fruhwirth2006finite}, a model  based on PDFs linear superposition, and as such each PDF $f_i$ is multiplied by a weight $\alpha_i$, i.e., $f_{mix}=\sum_i^N\alpha_if_i$, where N is the number of PDFs to combine. We then superimposed the PDFs of clusters of two, three, four, and up to N particles to construct a PDF which represents the magnitudes of $L_C$. 

\begin{figure}
	\centering
	\begin{subfigure}[h!]{0.48\textwidth}
		\includegraphics[scale=.4]{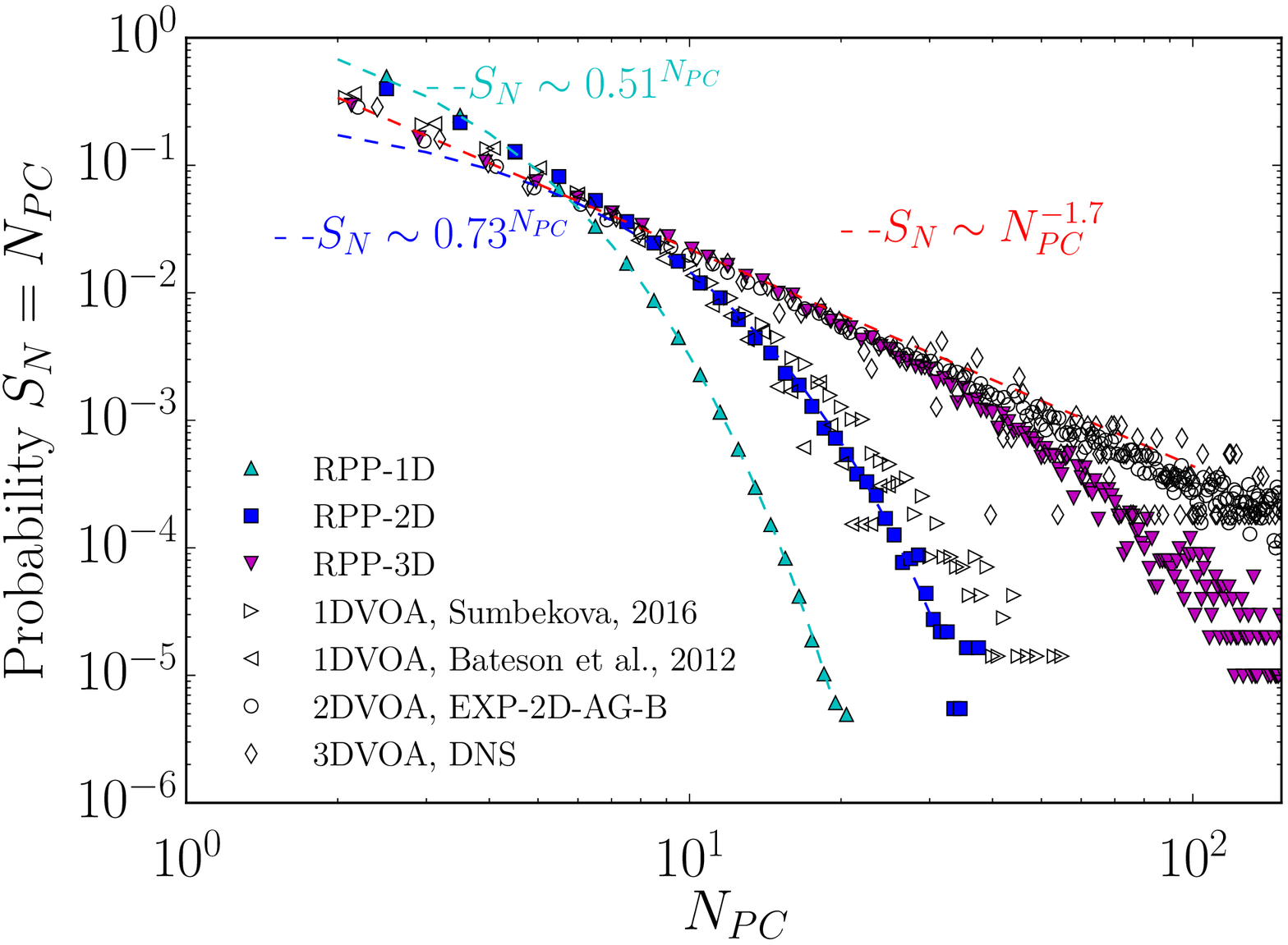}
		\caption{\label{hist-cl-2D-AA}}
	\end{subfigure}
	\quad
		\begin{subfigure}[h!]{0.48\textwidth}
		\includegraphics[scale=.4]{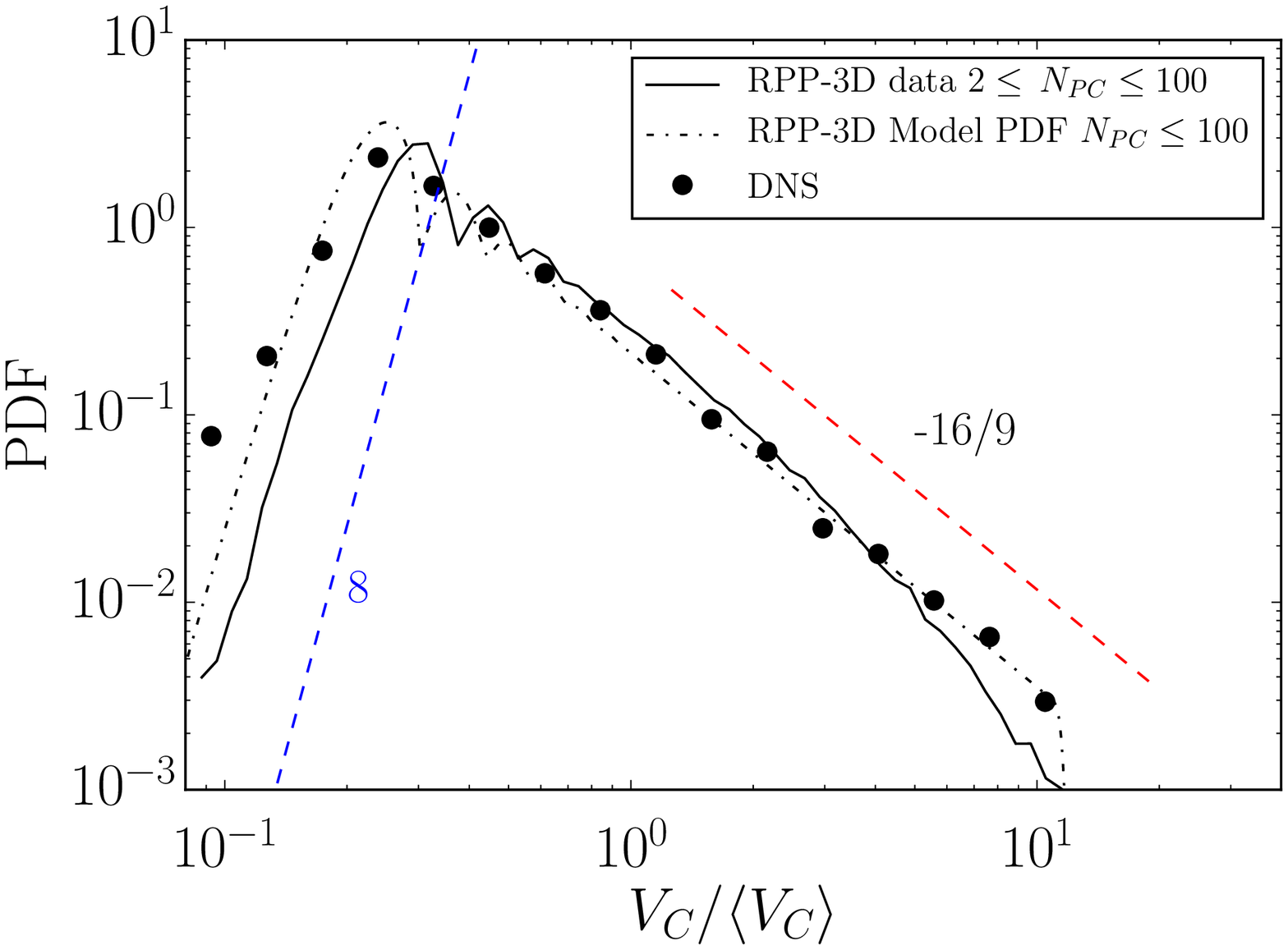}
		\caption{\label{fig-clu-3d}}
	\end{subfigure}

	\caption{a) Histogram for number of particles within a cluster the different datasets. Filled symbols represent random data, whereas open symbols represent the turbulent datasets. b) Probability distribution function of clusters volume $V_C/\langle V_C\rangle$ for 3D DNS data, synthetic data, and the PDF mixture model using the weights found in figure \ref{hist-cl-2D-AA}. The condition $\mathcal{V}\leq\mathcal{V}_{th}=0.62$ was employed for clustering computation.}

\end{figure}

The question was, however, how to choose the coefficients $\alpha_i$. A sensible approach is to propose that these weights are proportional to the probability of clusters having $i$ particles within the cluster collection. In other words, $\alpha_{N_{PC}}=$  counts of clusters of size $N_{PC}$ / the total number of clusters. We then obtained these weights by computing histograms ($S_N$) conditioned on the number of particles in a cluster; $N_{PC}$ (see figure \ref{hist-cl-2D-AA}, and equations \ref{eq:cpdf} - \ref{eq:nom}).

Our PDF mixture model based on these weights (see open symbols in figure \ref{hist-cl-2D-AA}), has good agreement with the clusters PDF found by applying the cluster identification algorithm to synthetic 1D RPP data (see figure \ref{fig-clu-rdn}), and therefore, it strongly supports the absence of a power law decay in the cluster size PDF coming from a 1D RPP distribution. 

Moreover, the construction of the histograms $S_N$ for turbulence-induced data reveals that $S_N$ histograms are power law distributed (see filled symbols in figure \ref{hist-cl-2D-AA}) up to certain extent for the data containing preferential concentration both in 1D and 2D, whereas they decay exponentially for randomly generated RPP data both in 1D and 2D. On the contrary, 3D randomly generated data still shows a power-law similar to the one present for the turbulent DNS dataset.

It therefore appears that the existence of a power law behavior in the clusters PDF is correlated to the functional dependency of the weights distribution ($S_N$) in any dimension (see figure \ref{hist-cl-2D-AA}). This observation then suggests that the individual PDFs ($f_i$) should attain a particular shape at increasing values of $N_{PC}$. Given that our model estimates the PDF of $L_C\vert_{N_{PC}}$ assuming a sum of independent random variables, i.e., $ L_C \vert_{N_{PC}}=X_1+ X_2+X_3+\ldots+X_{N_{PC}}$ (see in the appendix figure \ref{cl-nn30}, and equations \ref{eq:sumX} and \ref{eq:npcc}), we found that at increasing $N_{PC}$ this PDF tends to a Gaussian distribution, as stated by the central limit theorem \cite{ross1998first}. 

Although this theorem is valid for sums of independently distributed random variables, it has been rigorously proven \cite{deriggi2019central} (invoking strong-mixing conditions \cite{ibragimov1975note,bradley1981central}) that some consequences of it can extended to sums of correlated variables having limited normal or gamma distributions; $f_{\Gamma}(\mathcal{V},p,k)=k^p \mathcal{V}^{p-1}e^{-k\mathcal{V}}/\Gamma(k)$. Taking into account that; the Vorono\"{i} cells ($\mathcal{V}$) PDF (see table \ref{tab:fer}) are special cases, or numerical fits of the gamma distribution (for instance, in 1D $k=p=2$ \cite{kiang1966random,Ferenc2007}), and that sums of random correlated variables are implied when computing the clusters PDF, a central limit theorem \cite{deriggi2019central} should hold for the cluster algorithm of Monchaux et al. \cite{Monchaux2010}. We can then conclude that the histograms of $S_N$ indeed control shape of the cluster size PDF, and thereby, they provide a robust criterion to identify turbulence-driven clusters. 

From these previous results, three conclusions can be drawn. First, the weak power law regime seen in the right tail of the cluster size PDF in 1D is due to the limited extend of such behavior in their respective 1D $S_N$ histograms, as shown by the $S_N$ behavior computed from the experimental records of Sumbekova \cite{Sumbekova2016}, and Bateson et al.\cite{Bateson2012}. This behavior in the $S_N$ histograms is really robust and independent of the number of samples (for the PDI). For instance, the data of Sumbekova had  $10^6$ samples, whereas the data of Bateson et al. had $10^4$ samples. Hence, this weaker power law behavior is more likely to come from the loss of correlation inherent to the 1D measurements, and not due to insufficient statistical convergence.  

Second, the small `compact' clusters sizes, which are on the left tail of the PDF, have also a behavior close to a power law, i.e., $f_C(\mathcal{M}_C/\langle \mathcal{M}_C\rangle< 1)\approx(\mathcal{M}_C/\langle \mathcal{M}_C\rangle)^a$, where $\mathcal{M}$ stands for length $L$, area $A$ or volume $V$, and  $a$ is twice the absolute value of the algebraic exponent found in the base RPP Vorono\"{i} cell PDFs (see table \ref{tab:fer}), i.e., $a\approx2/5/8$ in 1D/2D/3D, respectively.

Third, the stronger power law behavior previously reported for the 2D/3D area/volume clusters PDF \cite{Obligado2014,baker2017coherent} can be alternatively explained by the much wider extent (compared to 1D) of such powerlaw behavior in the histograms of $S_N$ (figure \ref{hist-cl-2D-AA}). Our analysis is also in agreement with the findings of Uhlmann and collaborators \cite{uhlmann2014sedimentation,uhlmann2017clustering,chouippe2019influence} who report the presence of a power law for the clusters volume PDF (3DVOA) when the clustering algorithm is applied to a random Poisson distribution with no correlations at any scale. The latter result further supports the strong role that the $S_N$ histograms have on the clusters PDF despite the increased uncertainty of our mixture PDF model in higher dimensions due to the approximations involved in the base Vorono\"{i} PDFs in 2D and 3D (for details see \ref{aa:hd}, and table \ref{tab:fer}).

\begin{figure}
	\centering

		\begin{subfigure}[h!]{0.48\textwidth}
			\includegraphics[scale=.4]{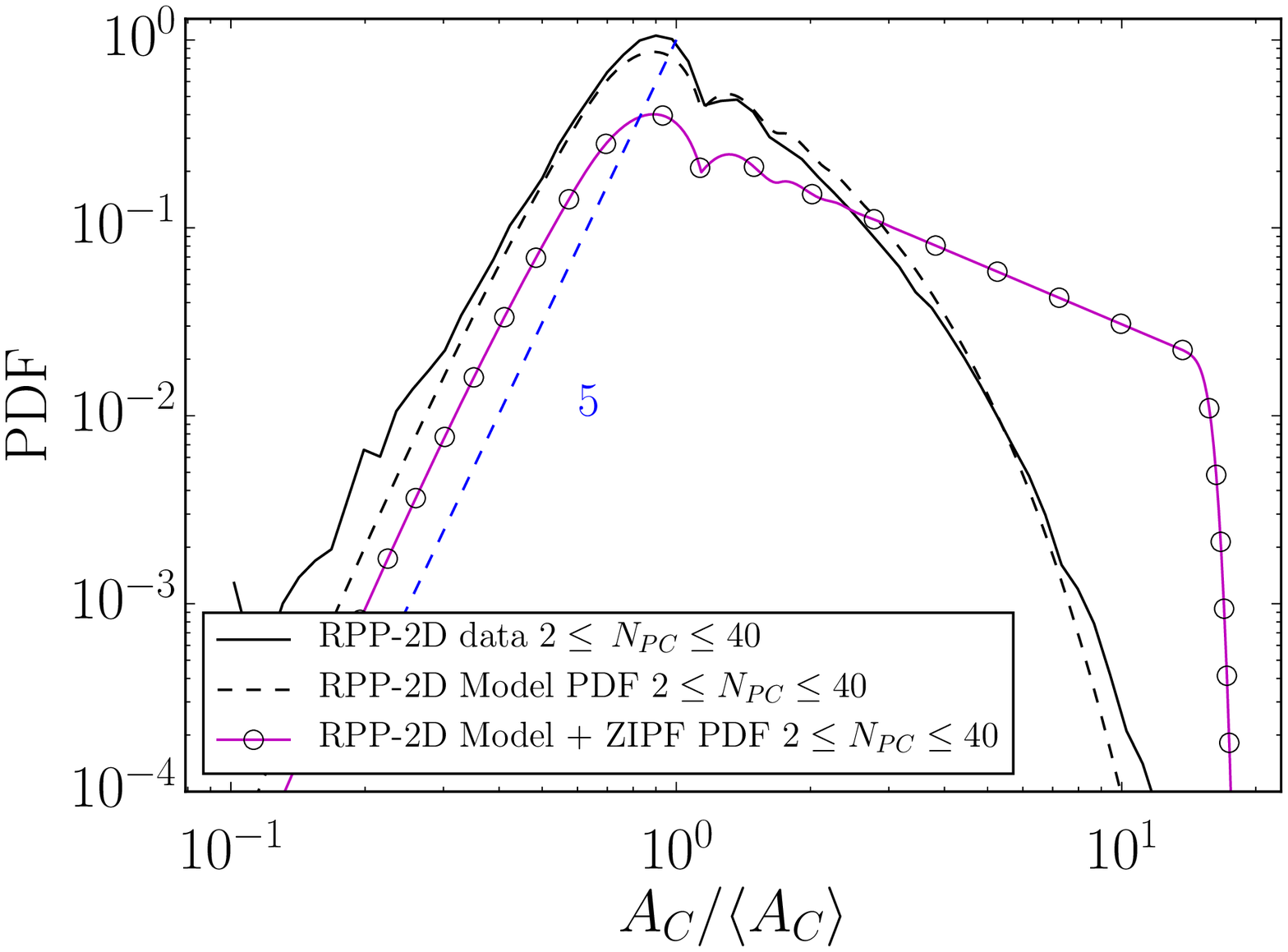}
			\caption{ \label{fig-clu-aa}}
		\end{subfigure}
	\quad
	\begin{subfigure}[h!]{0.48\textwidth}
		\includegraphics[scale=.4]{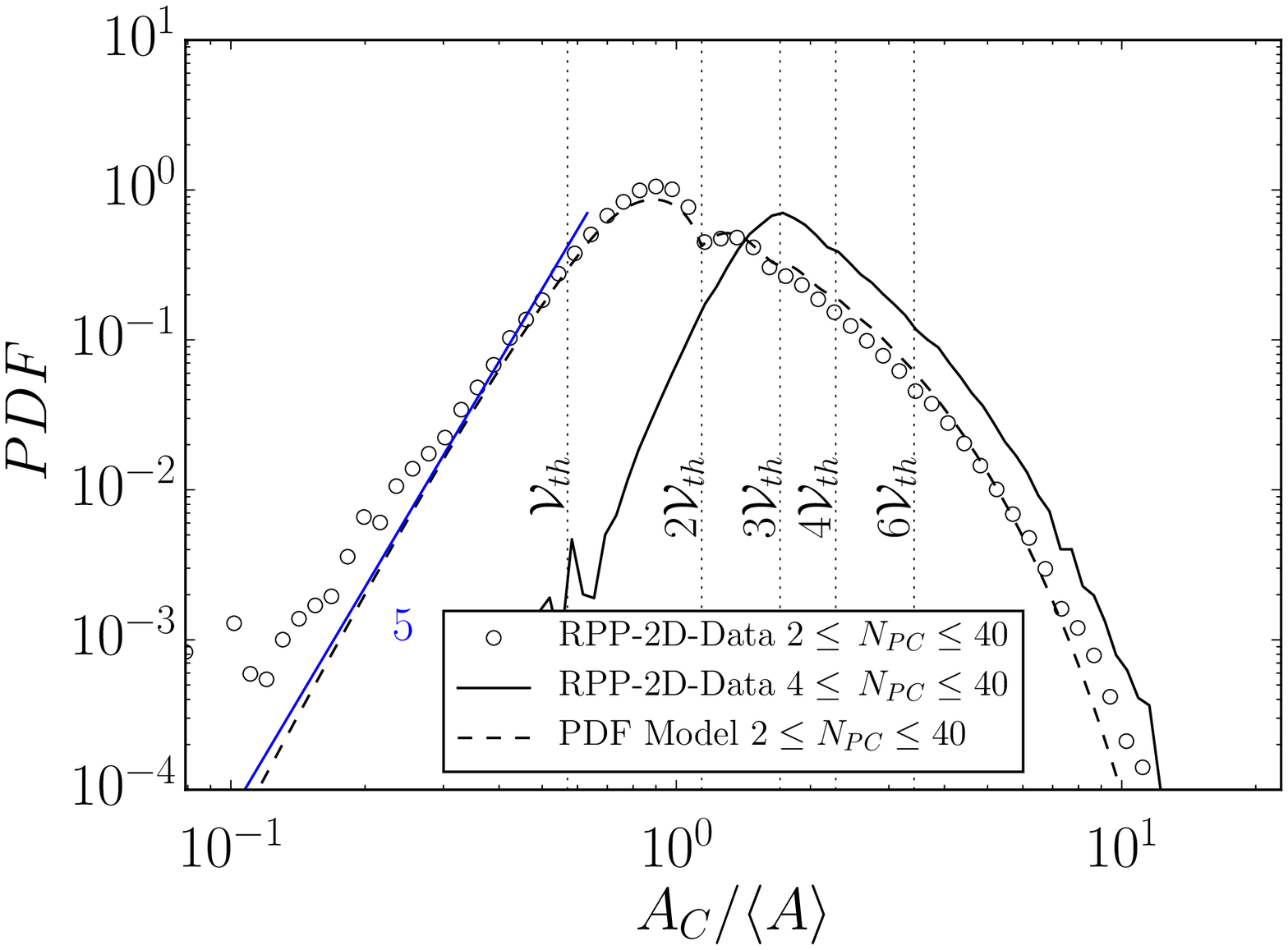}
		\caption{\label{fig-clu-ripp}}
	\end{subfigure}

	\caption{a)  Probability distribution function of cluster area $A_{C}/\langle A_C\rangle$ for a random uniform distribution, and the model proposed using the weights found in figure \ref{hist-cl-2D-AA}, and using weights from a ZIPF distribution $\alpha_i\propto N^{-1}_{PC}$. Indeed using the latter weights yields a strong power behavior. b) PDf of clusters area normalized by average Vorono\"{i} cell area $\langle A \rangle$ for the synthetic data, our model.  The presence of ripples within the analytical PDF seems to challenge the argument of Zamansky et al. \cite{zamansky2016turbulent} to cater for the 2DVOA edge effects.}
	
\end{figure}

Thus, it seems justified to use this model to analyze some properties, and possible biases found in 2D, and 3D Vorono\"{i} analyses, as our model for the PDFs of cluster areas/volumes have the same leading order behavior: with the absence/presence of power laws if the weights $S_N$ have exponential/power law distributions (see figures \ref{fig-clu-3d} , \ref{fig-clu-aa} and \ref{fig-clu-ripp}).

For instance, in the context of 2D Vorono\"{i} tessellations, an additional step has been recently suggested to suppress spurious edge effects \cite{zamansky2016turbulent,petersen2019experimental}. According to these studies, these edge effects cause ripples in the cluster areas PDFs, $A_C/\langle A_C\rangle$, and therefore, one has to require that all cells surrounding the detected clusters have a cell size below the threshold $\mathcal{V}_{th}$.

Although we recovered ripples in the 2D cluster area PDFs (figures \ref{fig-clu-aa} and \ref{fig-clu-ripp}), it is questionable that we could attribute the existence of these ripples to edge effects, as our analytical model (convolutions via Fourier transforms) should not be affected by them. From our model construction, it appears that these ripples occur at the boundary that mixes the individual PDFs, e.g., where the PDF for clusters containing 2 particles merges with the PDF for clusters containing 3 particles. Given that having clusters on these boundaries is less likely, it could explain such oscillations in the PDF. These ripples, however, disappear or become less `intense' for larger clusters sizes containing many particles. From our analysis, filtering out these ripples is equivalent to leaving out small, power law dependent clusters. 


\subsection{An approach to disentangle turbulence from random fluctuations}
\label{sc:appr}

Considering the results of the previous section, a complementary methodology to the classical algorithm \cite{Monchaux2010} is needed to discern turbulence-driven clusters from concentration fluctuations of random uncorrelated particles.

\begin{figure}
	\centering
	\begin{subfigure}[h!]{0.48\textwidth}
		\includegraphics[scale=.4]{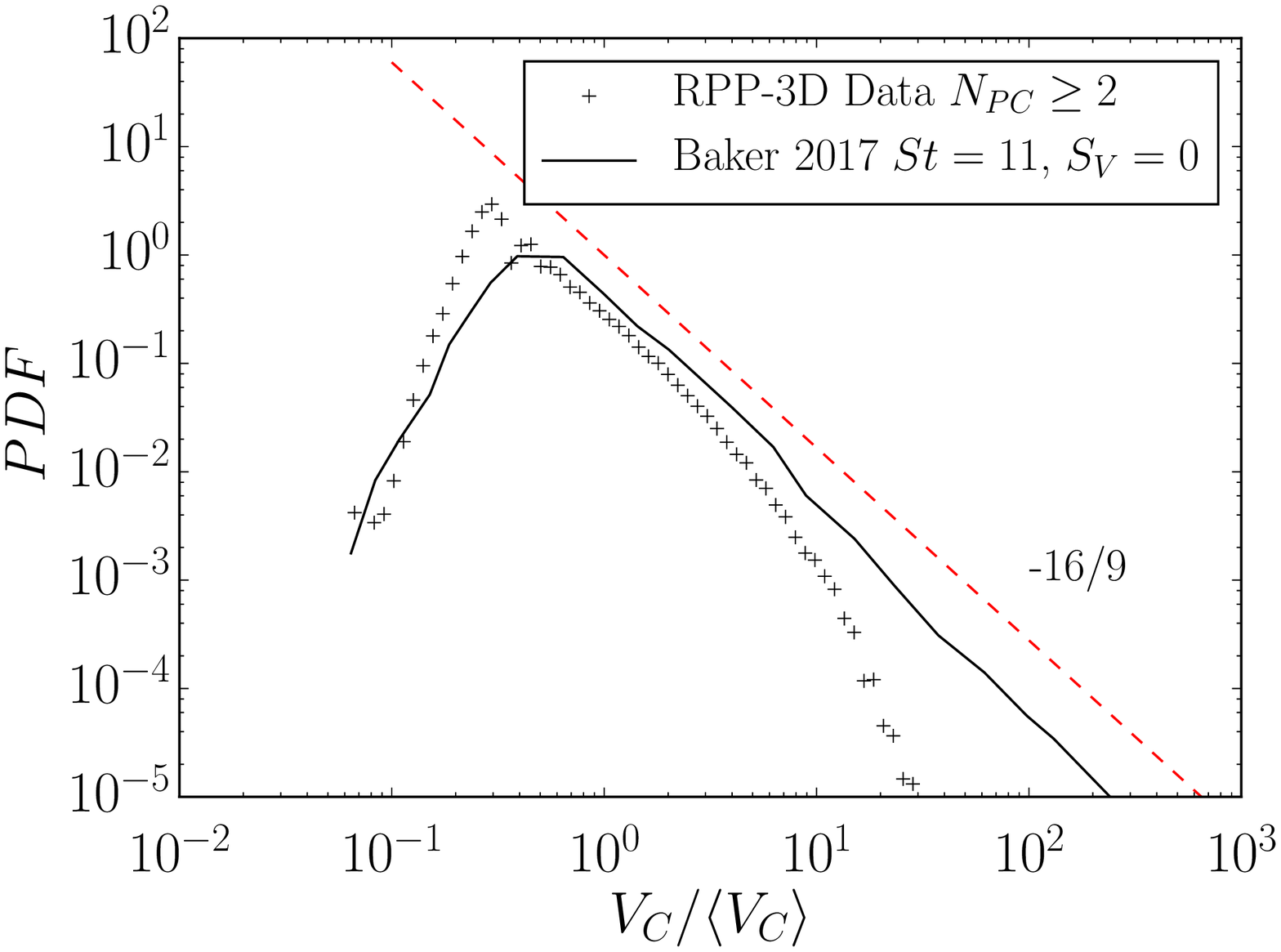}
		\caption{\label{fig-2d-clunpc-3D-cole}}
	\end{subfigure}
	\quad
	\begin{subfigure}[h!]{0.48\textwidth}
		\includegraphics[scale=.4]{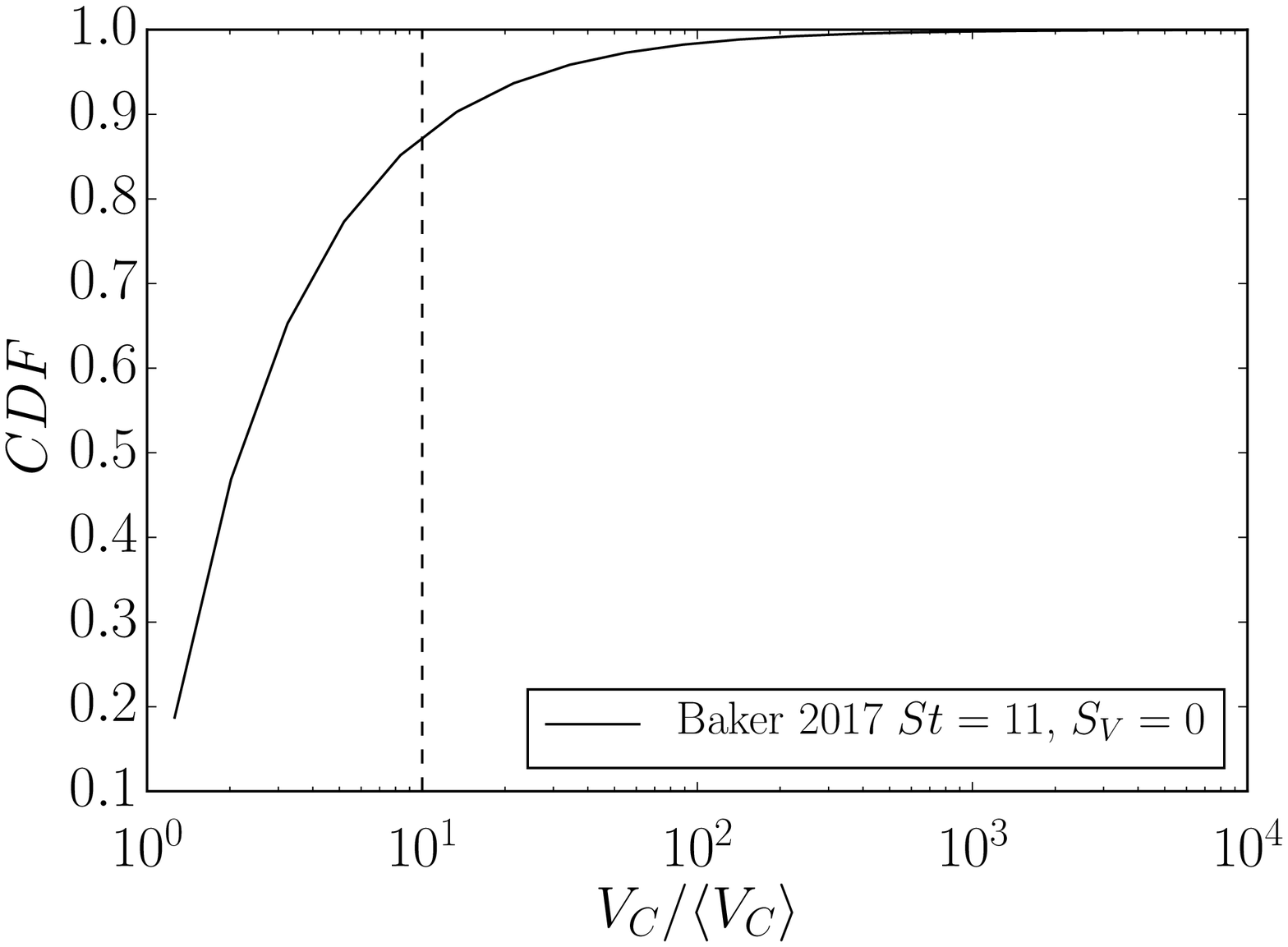}
		\caption{\label{fig-cdf-cole}}
	\end{subfigure}

	\caption{PDFs of 3DVOA (DNS data of Baker et al. \cite{baker2017coherent} for $St=11$, and $S_v=\tau_pg/u^\prime=0$) for normalized cluster volume $V_{C}/\langle V_C \rangle$. Baker et al. suggest that turbulence driven clusters are the ones for which for $V_C>\eta^3$ or $V_{C}/\langle V_C \rangle>1$. The figure shows that their criterion might fail (see figure \ref{fig-cdf-cole}) for almost the first decade of clusters (several points). b) CDF of 3DVOA (DNS data of Baker et al. \cite{baker2017coherent} for $St=11$, and $S_v=\tau_pg/u^\prime=0$) for normalized cluster volume $V_{C}/\langle V_C \rangle$. The vertical dashed line (\dashed) line represents up to the range where RPP-3D generated data follows a distinguishable power law (see figure \ref{fig-2d-clunpc-3D-cole}).}
\end{figure}

With this goal, we briefly recount proposed fixes found in the literature, and analyze their fitness to solve the mentioned problem. Baker et al. \cite{baker2017coherent} have proposed an amendment in the context of 3D Vorono\"{i} tessellations. Their underlying principle is that only the clusters lying on the right tail of  $V_{C}/\langle V_{C}\rangle$ (see figure \ref{fig-2d-clunpc-3D-cole}) should be considered as turbulence-driven `coherent' clusters. However, our data, and histogram analyses (figures \ref{hist-cl-2D-AA}, and \ref{fig-clu-3d}) challenge this argument, since the clustering algorithm applied to synthetic 3D RPP data still yields a self-similar behavior for the clusters volume distribution ($V_{C}/\langle V_{C}\rangle$). This behavior lasts for at least a decade, before transitioning into an exponential decaying behavior in agreement with the $S_N$ analysis. Thus, by following strictly this criterion, a large number of `random' (see figure \ref{fig-cdf-cole}) clusters could remain after the additional step proposed, and be counted as turbulence-induced clusters (see figures \ref{fig-2d-clunpc-3D-cole} and \ref{fig-cdf-cole}), defeating the purpose of such amendment.

An alternative test to distinguish randomness from turbulence-driven clusters has been proposed by Uhlmann and collaborators \cite{uhlmann2014sedimentation,chouippe2019influence}. It involves comparing the 3D clusters mean aspect ratio when clustering is present, as turbulence-driven clusters would have larger aspect ratio than randomly generated cells, due to gravity and enhanced particle settling. This criterion, however, could become less accurate as the clusters become less columnar.

On the other hand, the analyses presented in the previous section suggest a methodology based on segregated cluster probability $S_N$ (figure \ref{hist-cl-2D-AA}). According to these analyses, one expects that the right tail of the turbulence-driven cluster size distribution (in any dimension), conditioned on the number of particles in the cluster ($N_{PC}$) would preserve the powerlaw behavior at increasing thresholds of $N_{PC}$; whereas for random data this behavior would be eventually lost at increasing $N_{PC}$.

\begin{figure}

\begin{subfigure}[h!]{0.48\textwidth}
	\includegraphics[scale=.4]{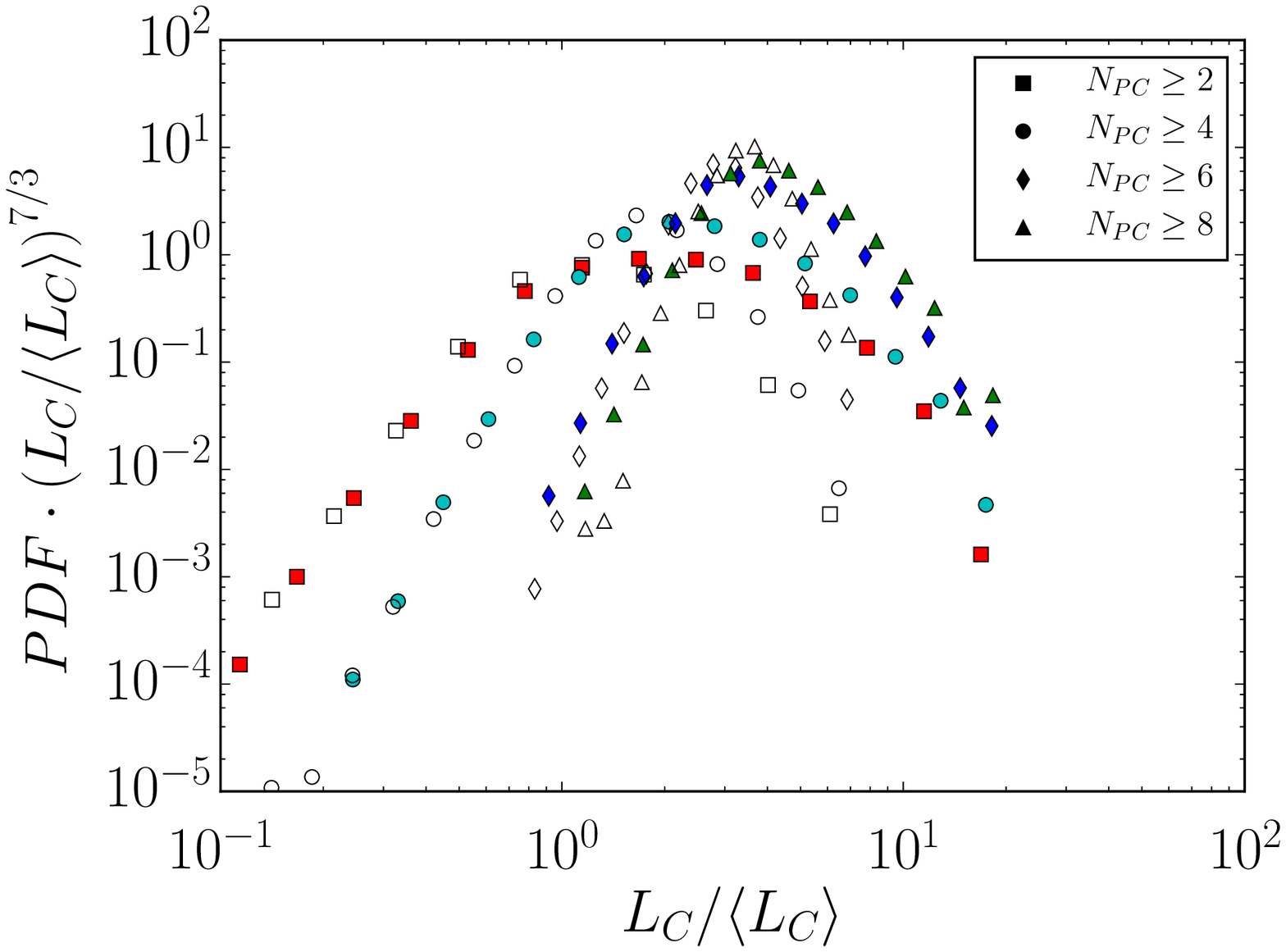}
	\caption{ \label{fig-1d-clunpc-rdn}}
	
\end{subfigure}	
	\quad
\begin{subfigure}[h!]{0.48\textwidth}
		\includegraphics[scale=.4]{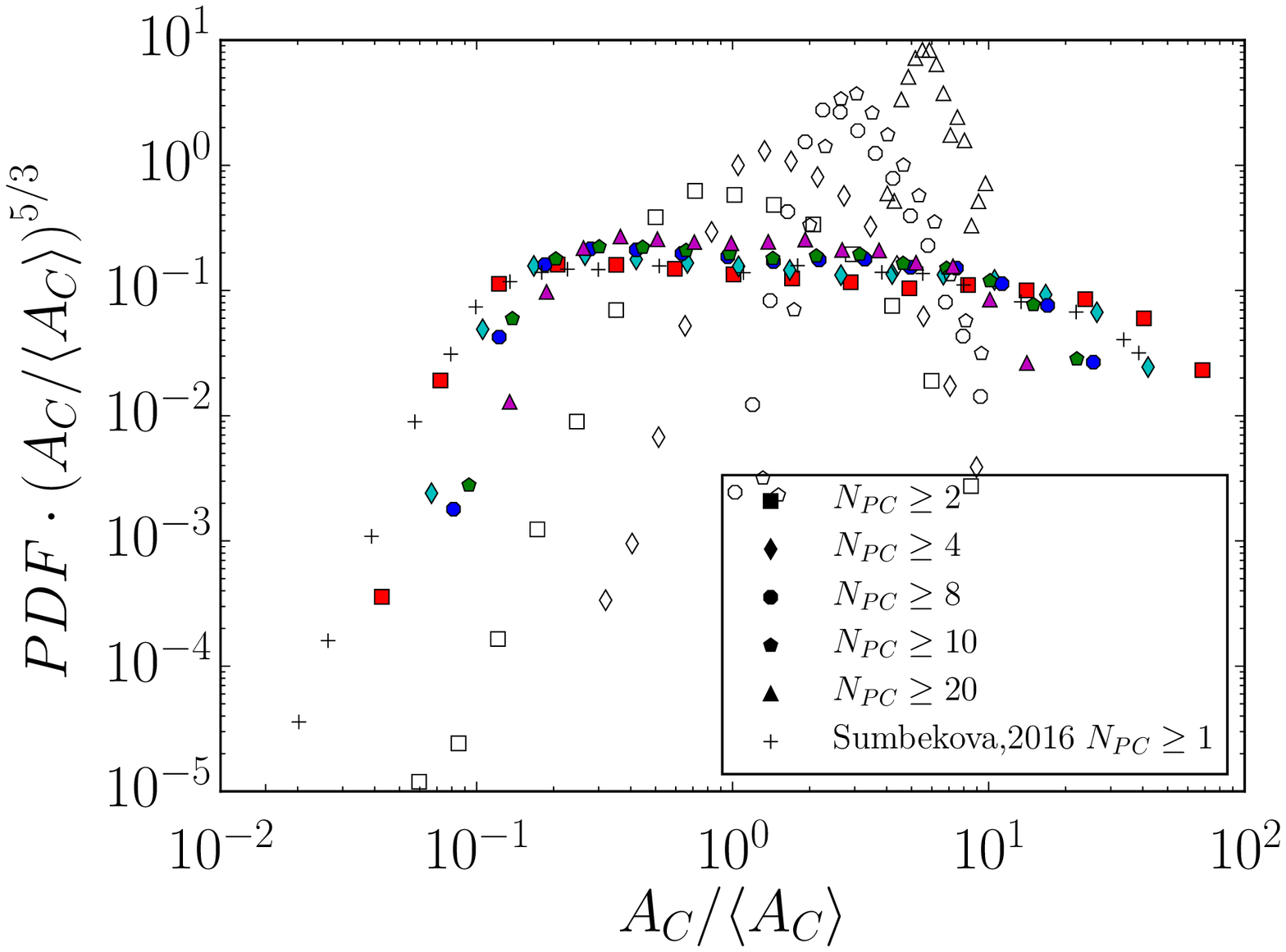}
		\caption{\label{fig-2d-clunpc}}
\end{subfigure}

\begin{subfigure}[h!]{0.48\textwidth}
		\includegraphics[scale=.4]{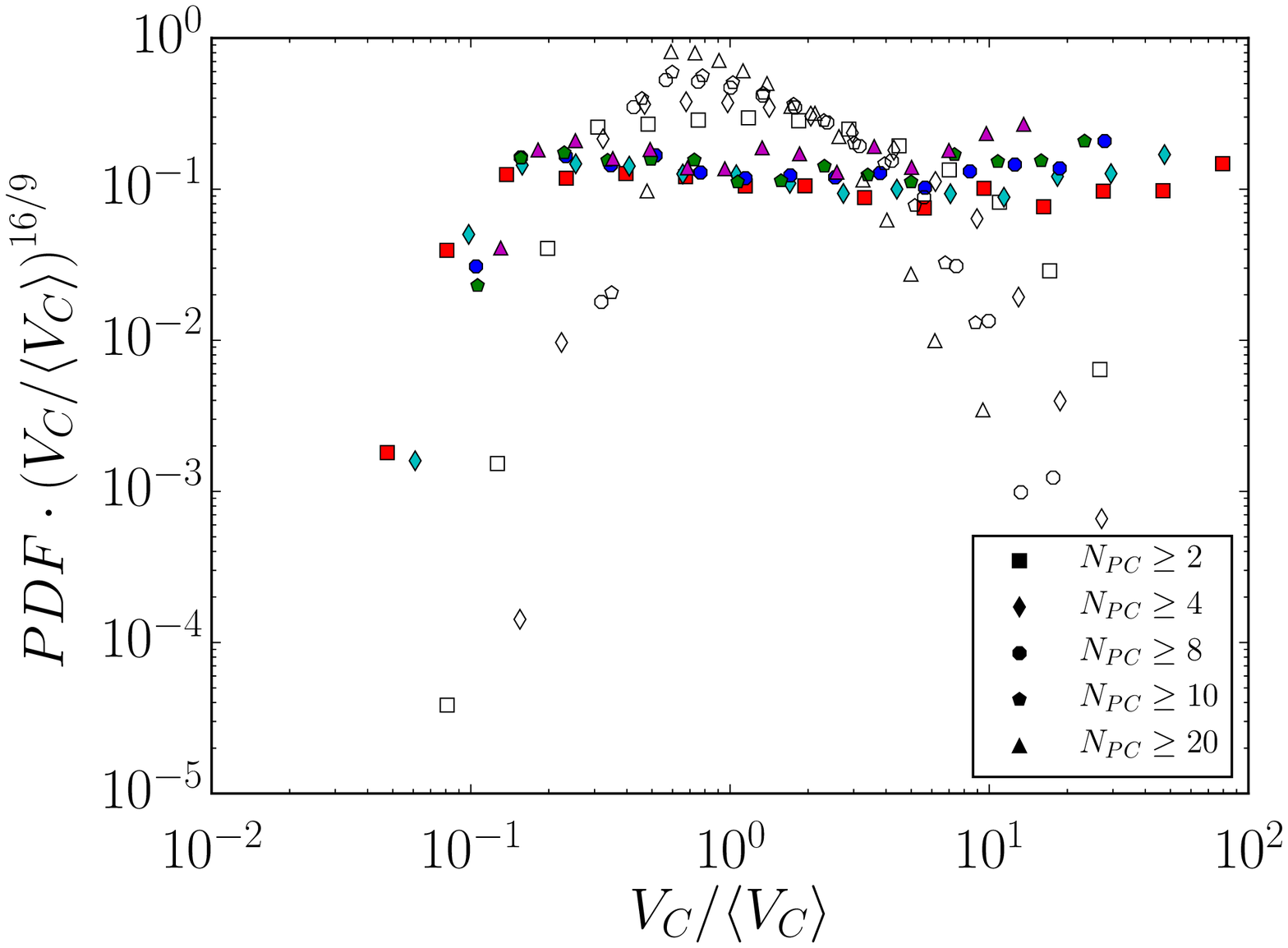}
		\caption{\label{fig-3d-clunpc-3D}}
\end{subfigure}

\caption{PDFs of characteristic cluster sizes compensated by the respective voids exponents given by the work of Goto and Vassilicos \cite{Goto2006} (7/3 and 5/3 for 1D and 2D in the order given), by Uhlmann and collaborators \cite{uhlmann2014sedimentation,uhlmann2017clustering} (16/9 for 3D). Filled symbols the come from the data from in table \ref{tab:par}, whereas open symbols are data coming from their respective RPP. a) PDFs of $L_{C}/\langle L_{C}\rangle$. Filled symbols are experimental data PDI from Sumbekova \cite{Sumbekova2016} and the 1D-RPP.
	b) PDFs of $A_{C}/\langle A_{C}\rangle$ for our 2D experimental data base, and the 2D-RPP c) PDFs of normalized cluster volume $V_{C}/\langle V_{C}\rangle$ for the DNS data, and the 3D-RPP.}

\end{figure}

For the 1D case, this segregation approach seems promising, as despite having a very weak power-law dependency (perhaps due to the limited amount of data from the records from Sumbekova \cite{Sumbekova2016}), the clustering containing data seems to preserve its power law behavior up to cluster of four or more points ($N_{PC}\leq4$) (see figure \ref{fig-1d-clunpc-rdn}).  

 In 2D, the segregation approach is very effective (figure \ref{fig-2d-clunpc}) as the right tail PDF power law behavior (figure \ref{hist-cl-2D-AA}) is conserved at increasing the values of $N_{PC}$. This is in agreement with the $S_N$ analysis (figure \ref{hist-cl-2D-AA}), and consistent with the mentioned conjecture that turbulence-driven clustering will preserve its power law behavior at increasing values of increasing $N_{PC}$. 

 In 3D, the clustering containing results apply to a larger extent their power law dependency, and the respective 3D random data eventually decays exponentially at around $N_{PC}>20$ (see figure \ref{fig-3d-clunpc-3D}), consistent with the transition found in $S_N$ (figure \ref{hist-cl-2D-AA}).

Thus, an analysis of the clusters histograms conditioned on the number of particles $N_{PC}$, is a promising complementary step to detect the presence of turbulence-driven clusters, as it seems capable of detecting and avoiding the possibles biases present in cluster identifying algorithms \cite{Monchaux2010, baker2017coherent} in any dimension.


\section{Clusters Concentration 1D}
\subsection{Average cluster concentration}

Knowledge of the particle local concentration is of the foremost importance in the study of particle-laden flows \cite{Balachandar2010}, and its potential applications. The easy access to local concentration maps is what gives Vorono\"{i} tessellations the upper hand with respect to different available methods \cite{Monchaux2010,Monchaux2012}.

However, taking into account the results from sections \ref{sc:prc}-\ref{sc-cs}
, the probe measuring volume will have an impact on the estimated local concentration values, when these tessellations are applied to 1D records.  In this section, we conduct a sensitivity analysis on the projected data (analogous to the one found in section \ref{sc-cs} for the average cluster size), and for the same data found in table \ref{tab:par}. We then define the cluster concentration as $C_C=N_{PC}/L_C$, where $N_{PC}$ is the number of particles inside the cluster, and $L_C$ is its length. The average concentration of the entire record is defined as $C_0=N_P/L_R$, where $N_P$ is the total number of particles detected over the total length recorded $L_R$, which is close to the inverse of the average Vorono\"{i} length \cite{Ferenc2007} $C_0\sim 1/\langle L \rangle =f$(MWS).

\begin{figure}[h!]
	\begin{center}
		\includegraphics[scale=.4]{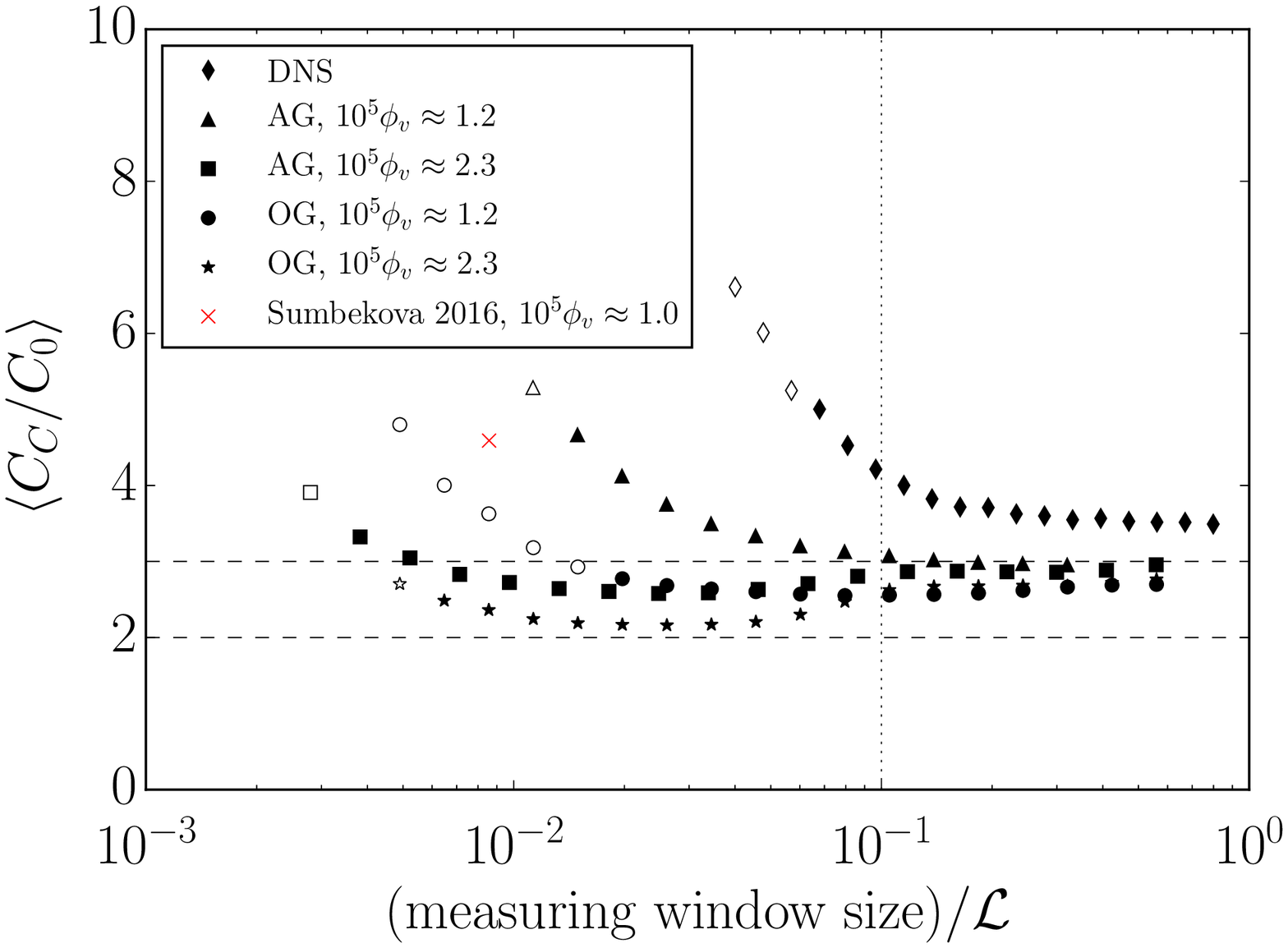}
		\caption{Average cluster concentration $\langle C_C / C_0 \rangle$ vs the measuring window scale for numerical and experimental data. Filled symbols refer to MWS for which $\sigma_\mathcal{V}/\sigma_{RPP}>1$, whereas blank symbols refer to $\sigma_\mathcal{V}/\sigma_{RPP}<1$ (figure \ref{fig-std-2dsc-L}). The X refers to the PDI data from Sumbekova\cite{Sumbekova2016a} $\langle C_C / C_0 \rangle\sim 4.56$. The larger the number of points projected, the closer the cluster concentration to the value of 3. \label{fig-avg-AA}}
	\end{center}
\end{figure}

The analysis by varying windows shows that the average cluster concentration value $\langle C_C / C_0 \rangle $ has a non-monotonic behavior which depends on the measuring window size, and the bulk liquid fraction $\phi_v$ (although only one dataset is shown in figure \ref{fig-avg-AA} the remaining data sets -not shown here- exhibited the same behavior). There is also a transition region similar to the one found in the normalized average cluster size  $\langle L_C\rangle/\eta$ (see figure \ref{fig-cll}), and after this region the concentration tends to a value close to 3.

For the transition region, the data points with a measuring size of order $\eta$ are in good agreement with previous 1D and 2D experimental data, i.e., $\langle C_C / C_0 \rangle \in [2.0-5.0]$ \cite{Monchaux2010,Sumbekova2016,obligado2019study}. Although a comparison between the 1D and the 2D average cluster concentration values $\langle C_C / C_0 \rangle $  was not conclusive due to the uncertainty in the lower to higher dimension extrapolation \cite{Spinewine2003}, it seems that our approach also captures the leading order evolution of the average cluster concentration. 

\textcolor{black}{
The latter supports again that under similar experimental conditions an instrument with  with measuring window size of order $\eta$, is able to recover global average clustering parameters which are in agreement with previous data captured with higher dimensional techniques.}

On the other hand, at increasing MWSs values, the cluster concentration approaches asymptotically to $\langle C_C / C_0 \rangle \approx 3.0$; cluster density varies proportionally with the MWS in agreement with arguments of section \ref{sc-cs}.  It is, however, important to notice that $\langle C_C / C_0 \rangle \approx 3.0$ is very close to the value attained when the same algorithm is applied to the denser regions of RPP distributions in any dimension. We can estimate this concentration analytically for the RPP distribution, if we take account that $C_0/C\approx \langle L\rangle\vert_{0}^{\mathcal{V}_{th}}=K\int_{0}^{\mathcal{V}_{th}}\mathcal{V}^2e^{-2\mathcal{V}}d\mathcal{V}$ ($K$ is a normalization constant see equation \ref{eq:pdf}) for $\mathcal{V}_{th}=0.55$, $C_0/C\approx \langle L\rangle\vert_{0}^{\mathcal{V}_{th}}\approx 0.332$, or $C/C_0\approx 3$. Given the similarity between the RPP, and turbulence-driven values, care has to be taken when drawing conclusions based on average cluster concentration.

\subsection{Concentration PDFs}

\begin{figure}
\centering

\begin{subfigure}[h!]{0.48\textwidth}
		\includegraphics[scale=.48]{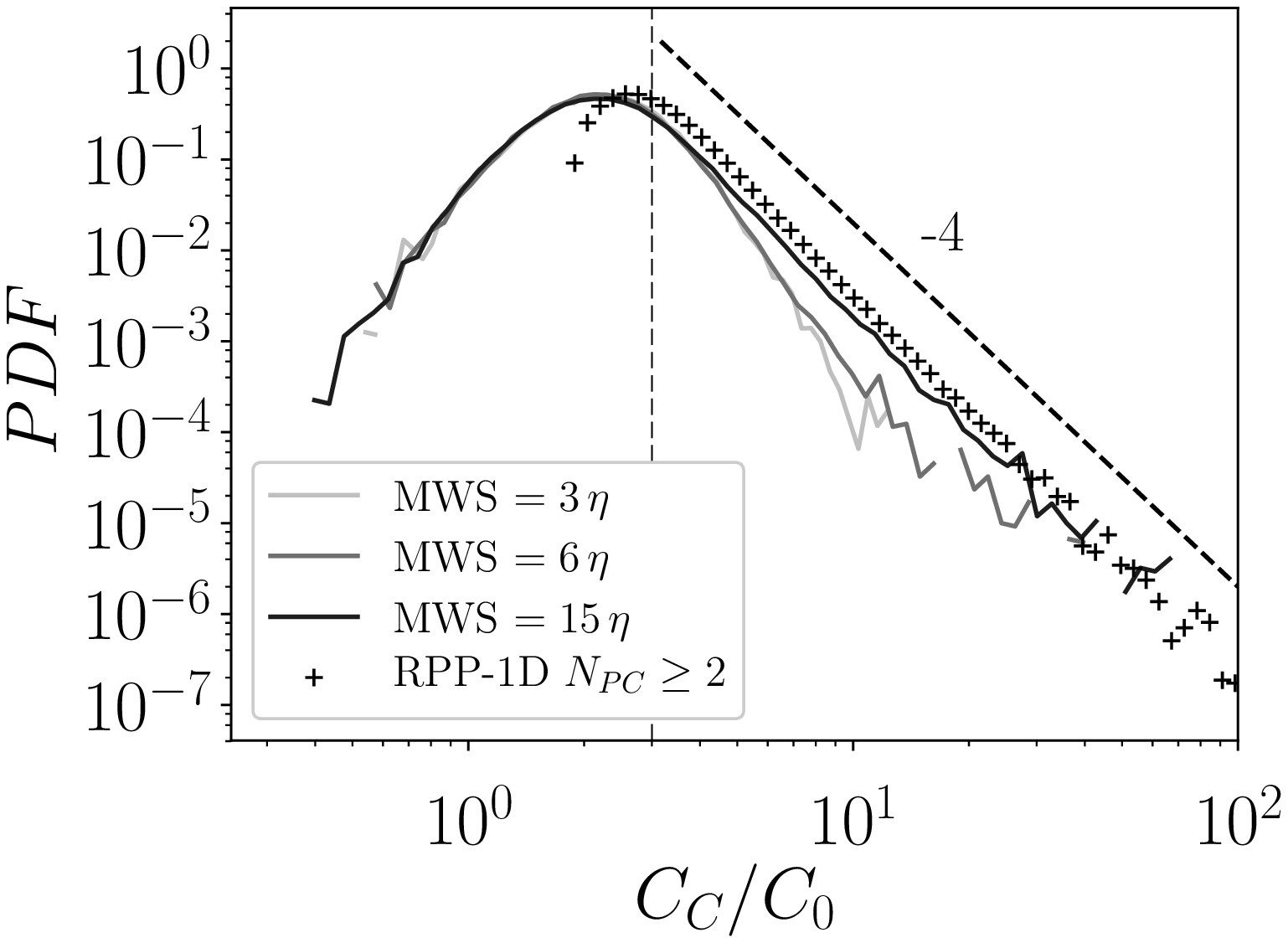}
		\caption{\label{fig-pdf-cc-AA-AG}}
\end{subfigure}
\begin{subfigure}[h!]{0.48\textwidth}
		\includegraphics[scale=.4]{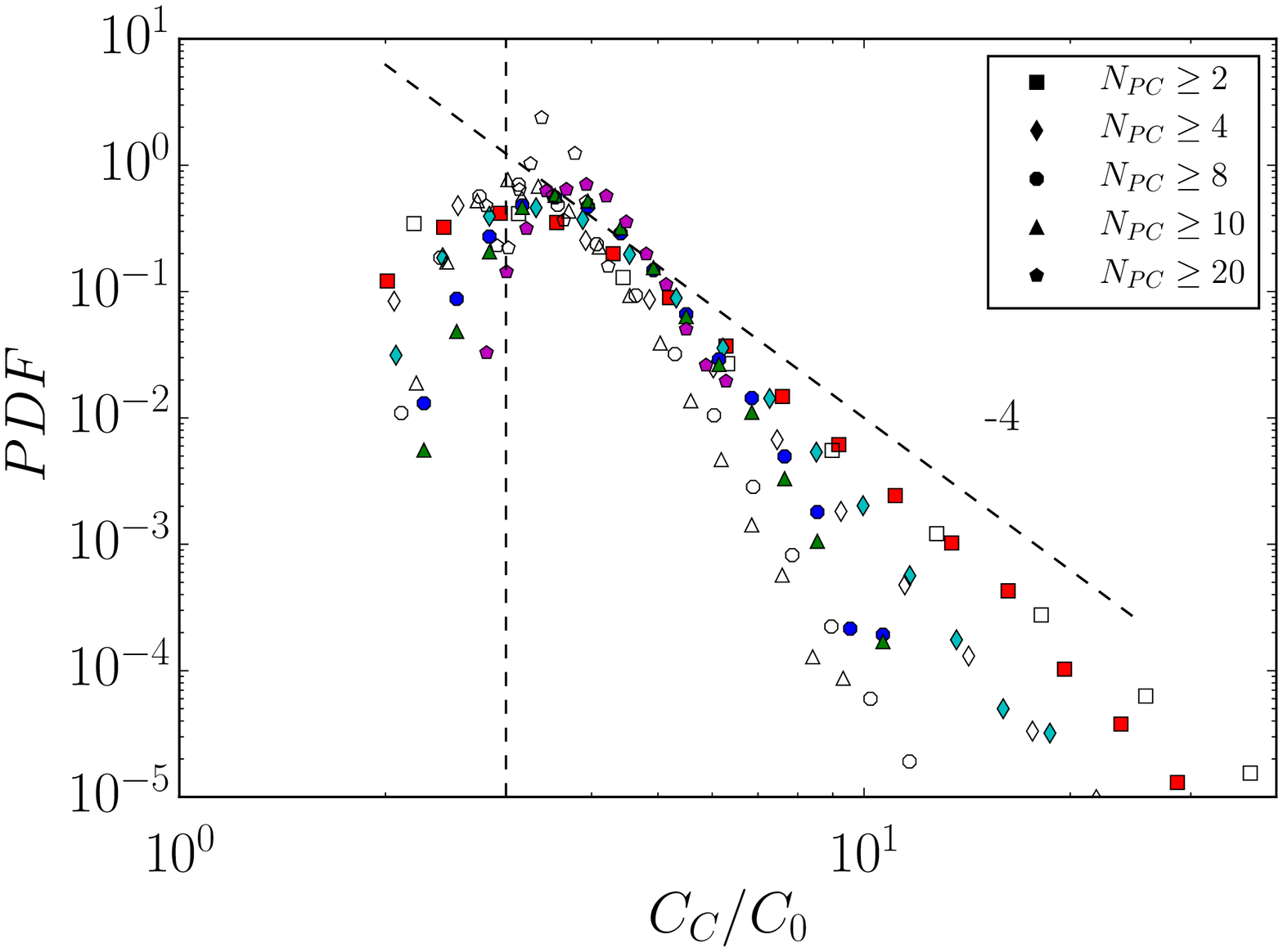}
		\caption{\label{CNORM-SH}}

\end{subfigure}

\caption{ a) PDF of the cluster concentration over global concentration $C_C / C_0 $. For experimental sampled data (EXP-2D-AG-B) ($2D_{EXP} \rightarrow 1D_\perp$), and different measuring window sizes (MWS). b) PDF of the cluster concentration over global concentration $C_C / C_0 $. Solid symbols are experimental PDI data \cite{Sumbekova2016a}, whereas open symbols are from RPP-1D data. The vertical (\dashed) line represents the asymptotic value $C_C/C_0 \approx 3.0$ found in figure \ref{fig-avg-AA}. It can be seen that large cluster concentrations have higher probability in clustering containing data than in random data.}
\end{figure}

We now explore the concentration PDFs, and their sensitivity to variations with the measuring window size. Similar to the clusters PDF, the 1D concentration PDFs of synthetic random and clustering containing data have a close resemblance (figure \ref{fig-pdf-cc-AA-AG}).

The concentration PDF also follows a $-4$ exponent, as reported by Sumbekova via PDI measurements \cite{Sumbekova2016a}. This exponent, however, is a trivial result of the unidimensional tessellation, as it can be analytically estimated; given that $C_C\sim L^{-1}_C$ , so that PDF$_{C_C}$($C_C$) $\sim$ PDF$_{L_C}$($1/C_C$)$\times C^{-2}_C$ by the chain rule \cite{hogg2005introduction}. Then, for the region $L_C/\langle L\rangle \ll1$, $C_C\gg1$, where the cluster size PDF$_{L_C}$ exhibits a power law with exponent of $2$ (figure \ref{fig-clu-rdn}), it follows that PDF$_{C_C}$($C_C$)$\sim L_C^{4}\sim C_C^{-4}$, and thereby, the -4 exponent.

Thus, to gain more insight into the effects of turbulence on $C_C$, we conditioned the clusters based on their number of particles, as proposed in section \ref{sc:appr}. Although mild differences are found between clustering, and random generated data at very small $N_{PC}$, the preferential concentration data shows that it is almost an order of magnitude more likely to find very dense regions $C_C/C_0>4$ than in the RPP (see figure \ref{CNORM-SH}) when $N_{PC}>8$. Previous experiments have shown that particles belonging to `compact' (very dense) clusters exhibit peculiar properties, for instance, enhanced settling velocity due to collective effects \cite{Aliseda2002, huck2018role}. Given these results, it is questionable to neglect those `compact' clusters  found at the left region of the clusters PDFs, as proposed by the amendment of Baker et al. \cite{baker2017coherent} also discussed in section \ref{sc:appr}. 

As a final observation, it seems advisable that future experiments should acquire very long records in order to disentangle, and correctly study statistics conditioned on local concentration as 1D measurements are unable to register enough turbulence-driven clusters events. For instance, the PDI data of Sumbekova \cite{Sumbekova2016}, contained in its entirety $10^6$ samples, which yielded $10^4$ clusters with $N_{PC}\geq2$, and $10^3$ clusters for $N_{PC}\geq10$, which undeniably impacts the convergence of any conditioned statistic.


%




\section{Conclusions}

Unidimensional Vorono\"{i} tessellations  to analyse preferential concentration provide results consistent with the mean values of cluster size and concentration reported in previous studies under the same experimental conditions. However, quantifying preferential concentration by means of this analysis has some biases that need to be considered.

It is clear that if unidimensional measurements via 1DVOA recover evidence of particle clustering, i.e., $\sigma_\mathcal{V}/ \sigma_{RPP}>1$ following the criteria of Monchaux et al.\cite{Monchaux2010}, this evidence is reliable as random distributions cannot create spurious traces of clustering given adequate statistical convergence.

On the contrary, if 1DVOA yields $\sigma_\mathcal{V}/ \sigma_{RPP}\approx 1 $, it cannot be directly concluded that preferential concentration is absent within the flow. 

The origin of this effect has been tracked to the measuring window size (MWS):  if it is too small or too wide, it might not be able to detect preferential concentration present within the turbulent flow by means of the unidimensional Vorono\"{i} Analysis (1DVOA). Thus, MWS plays an important role on preferential concentration measurements that needs to be understood and taken into account. 

\textcolor{black}{To capture preferential concentration with 1DVOA, MWS should be a fraction of the expected cluster length, $\langle L_C\rangle$. In the experimental conditions reported here, our results suggest the range of MWS that retrieve preferential concentration may have as lower bound the Kolmogorov length scale ($\eta$). }

This guidelines are justified by our sensitivity analysis, which shows that the maximum degree of clustering (the magnitude of $\sigma_\mathcal{V}$) is found at MWS$_\star\sim\mathcal{L}/10$, where $\mathcal{L}$ is the integral length scale of the carrier phase. In general, this expression seems to follow MWS$_\star\sim \langle L_C\rangle$, which combined with MWS$_\star\sim\mathcal{L}/10$, yields that this window should be of order $\eta$ consistent with previously reported values of $L_C=\mathcal{O}(10-20)\eta$ under similar conditions.

\textcolor{black}{
Hence, a MWS within the latter range may be able to recover meaningful values of the average cluster linear size $\langle L_{C}/\eta \rangle$, as well as the average cluster concentration $\langle C_C/C_0 \rangle $.}

However, it seems that these thresholds (in general) could depend on specific experimental conditions, or DNS simulation conditions, and therefore an iterative procedure for varying the MWS should be put in place if evidence of preferential concentration is not recovered at the first try by unidimensional analysis. This conclusion is supported by the 1DVOA performed on the numerical projections of 3D numerical, and 2D experimental data ($3D_{DNS}\rightarrow1D_{\perp}$, $2D_{EXP}\rightarrow1D_{\perp}$, respectively), and previously reported quasi-unidimensional PDI data.

A pitfall of the 1DVOA is that the `raw' cluster linear size $L_C/\langle L_C \rangle $ and cluster concentration $C_C / C_0 $ probability density functions (PDFs) might not be insightful for characterizing preferential concentration, as the loss of information inherent to the 1D projections in quasi-unidimensional experimental methods, e.g., PDI, or optical probes, weakens the correlations present within data, and yields as consequence the disapperance of a power law behavior in the right tail of the cluster size PDF.

We develop a simple theoretical model for cluster PDFs from random distributions, which ruled out the existence of a power law within the resulting PDF.  However, given the urgent need of characterizing these PDFs, we proposed an alternative filtering approach  to disentangle randomness from turbulence clustering. This approach conditions the cluster size PDF by the number of particles in a cluster, $N_{PC}$, and is based on the premise that if preferential concentration is present in the data, the right tail of the clusters PDF will conserve its power law dependence, as in respective clusters histogram $S_N$. This could have deep implications when analyzing the impact of collective effects on particles settling velocity via quasi-unidimensional measuring techniques. 

Our data suggests that the 2D and 3D Voronoi tessellations (2DVOA/3DVOA) are much more robust to characterize the cluster size PDF than its 1D counterpart. In fact, 2DVOA is very robust and its biases are minimal, as previously demonstrated \cite{Romain2012}. 

3DVOA, however, presents biases that a recently proposed cutoff criterion \cite{baker2017coherent} does not consider. The results presented for the classical cluster detection algorithm \cite{Monchaux2010} applied to 3D random synthetic data, which does not contain correlations at any scale, yielded the presence a power law on the right tail of the cluster size PDF, as reported elsewhere \cite{uhlmann2014sedimentation,uhlmann2017clustering}. In 3D, our filtering approach seems promising, as it allows to clearly determine whether the clusters data is coming from a random, or turbulent origin. In the latter case our results confirm that a powerlaw is preserved for a larger extent when compared to the former case. 

\section{Acknowledgments}

The authors would like also to thank A2PS for all the material, and support provided during the study of the optical probe. We thank Markus Uhlmann and Agatha Chouippe from KIT for fruitful discussions, and  Laure Vignal for her help and insight regarding the experimental setup. The numerical (DNS) Lagrangian data set was provided  by the TurBase \cite{benzi2017turbase} repository under the project EuHIT: a European High performance Infrastructures in Turbulence (Grant agreement $\#$ 312778).  Our work has been partially supported by the LabEx Tec21 (Investissements d’Avenir Grant
Agreement No. ANR-11-LABX-0030), and by the ANR Project No. ANR-15-IDEX-02.

\appendix
\section{Clusters PDF Model RPP Distribution}
\label{sc:Ap}

In order to model the normalized cluster size PDF for $L_C/\langle L_C \rangle $ coming from randomly generated data (RPP distribution, see table \ref{tab:fer}). The following assumptions were made:

\begin{itemize}
\item The normalized cluster size PDF for a cluster population with an arbitrary number of cluster points  $L_C/\langle L \rangle \vert_{N_{PC}} $, where $N_{PC}\geq2$ is the number of points inside the cluster, is equal to the sum $N_{PC}$ of the independent and identically distributed \cite{johnson1994continuous} variables $X_i$, with $i=1,2,3,\ldots,N_{PC}$;

\begin{equation}
\frac{L_C \vert_{N_{PC}}}{\langle L \rangle}=X_1+ X_2+X_3+\ldots+X_{N_{PC}}
\label{eq:sumX}
\end{equation}

The random variable $X_i$ is distributed as:
\begin{equation}
 f(\mathcal{V})=K_f
\begin{cases}
	\mathcal{V}e^{-2\mathcal{V}} & \text{if $0<\mathcal{V}<\mathcal{V}_{th}$} \\
	0 & \text{otherwise} 
\label{eq:pdf}
\end{cases}
\end{equation}

\begin{equation}
 \langle\mathcal{V}\rangle\vert_{0<\mathcal{V}<\mathcal{V}_{th}}=\frac{e^{2\mathcal{V}_{th}}-2\mathcal{V}_{th}^2-2\mathcal{V}_{th}-1}{e^{2\mathcal{V}_{th}}-2\mathcal{V}_{th}-1}
\end{equation}

\begin{equation}
\frac{\langle L_C \vert_{N_{PC}}\rangle}{\langle L\rangle}=N_{PC}\langle\mathcal{V}\rangle\vert_{0<\mathcal{V}<\mathcal{V}_{th}}
\label{eq:NAVG}
\end{equation}

where $\mathcal{V}$ is the normalized cell size, $K_f=0.25\big(1-(2\mathcal{V}_{th}+1)$exp$(-2\mathcal{V}_{th})\big)$ a constant that accounts for the normalization of the PDF, namely,\\ $\int^{\mathcal{V}_{th}}_0 f(\mathcal{V}) d\mathcal{V}=1$. $f$ is the theoretical model PDF for $\mathcal{V}$  proposed by Ferenc et N\'{e}da \cite{Ferenc2007} but with its domain bounded by $\mathcal{V}_{th}$, which is the threshold to compute the clusters. 

The sum of $N_{PC}$ independent variables has a PDF equal to the convolution of their respective individual PDFs  \cite{hogg2005introduction}, e.g., $N_{PC}=2$ (the simplest case):

\begin{equation}
f_2(Z)=\int^\infty_{-\infty} f_X(X)f_Y(Z-X)dX
\end{equation}

with $Z=X+Y$ (these are dummy random variables, it is equally valid $Z=X_1+X_2$), considering the support of components distribution, a traditional technique consist in dividing the range of the new random variable in two:

\begin{equation}
f_2(Z) =K_2
\begin{cases}
\int^{Z}_0Xe^{-2X}(Z-X)e^{2(X-Z)}dX& 0<Z\leq \mathcal{V}_{th} \\
\int^{\mathcal{V}_{th}}_{Z-\mathcal{V}_{th}}Xe^{-2X}(Z-X)e^{2(X-Z)}dX& \mathcal{V}_{th}<Z\leq 2\mathcal{V}_{th} 
\end{cases}
\end{equation}

Where $K_2$ is a normalizing constant. As the original variables had a support from $0<Z\leq \mathcal{V}_{th}$ is straight forward to see that the new support is $0<Z\leq 2\mathcal{V}_{th}$, further simplifying:

\begin{equation}
f_2(Z) =K_2e^{-2Z}
\begin{cases}
\int^{Z}_0XZ-X^2dX& 0<Z\leq \mathcal{V}_{th} \\
\int^{\mathcal{V}_{th}}_{Z-\mathcal{V}_{th}}XZ-X^2dX& \mathcal{V}_{th}<Z\leq 2\mathcal{V}_{th} 
\end{cases}
\end{equation}

After carrying on the integration:
\begin{equation}
f_2(Z) =K_2e^{-2Z}
\begin{cases}
Z^3& 0<Z\leq \mathcal{V}_{th} \\
	6Z \mathcal{V}_{th}^2-Z^3-4 \mathcal{V}_{th}^3&  \mathcal{V}_{th} <Z\leq 2\mathcal{V}_{th} 
\end{cases}
\label{eq:nn2}
\end{equation}

The figure \ref{cl-nn2} illustrates that there is good agreement with the right tail of the distribution between numerically generated data, and the model here proposed. The disagreement in the left tail comes from the nature of this assumption, as we ignore the influence of the neighbours in the construction of the Vorono\"{i} cells. However, it seems that this approximation is satisfactory for the right tail of the distribution.

\begin{figure}
	\centering
\begin{subfigure}[h!]{0.48\textwidth}
	\includegraphics[scale=.4]{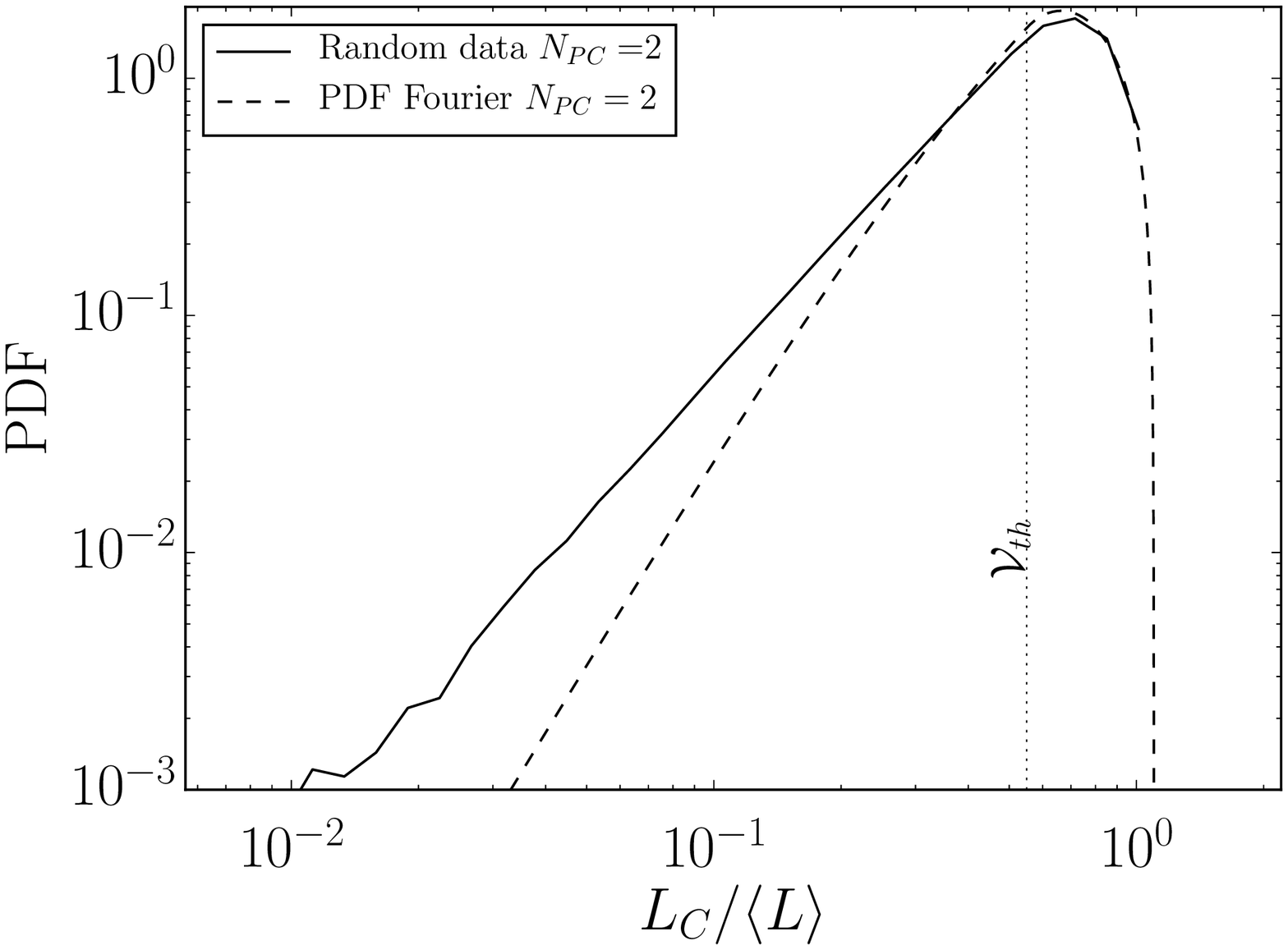}
	\caption{\label{cl-nn2}}
\end{subfigure}
\quad
\begin{subfigure}[h!]{0.48\textwidth}
		\includegraphics[scale=.4]{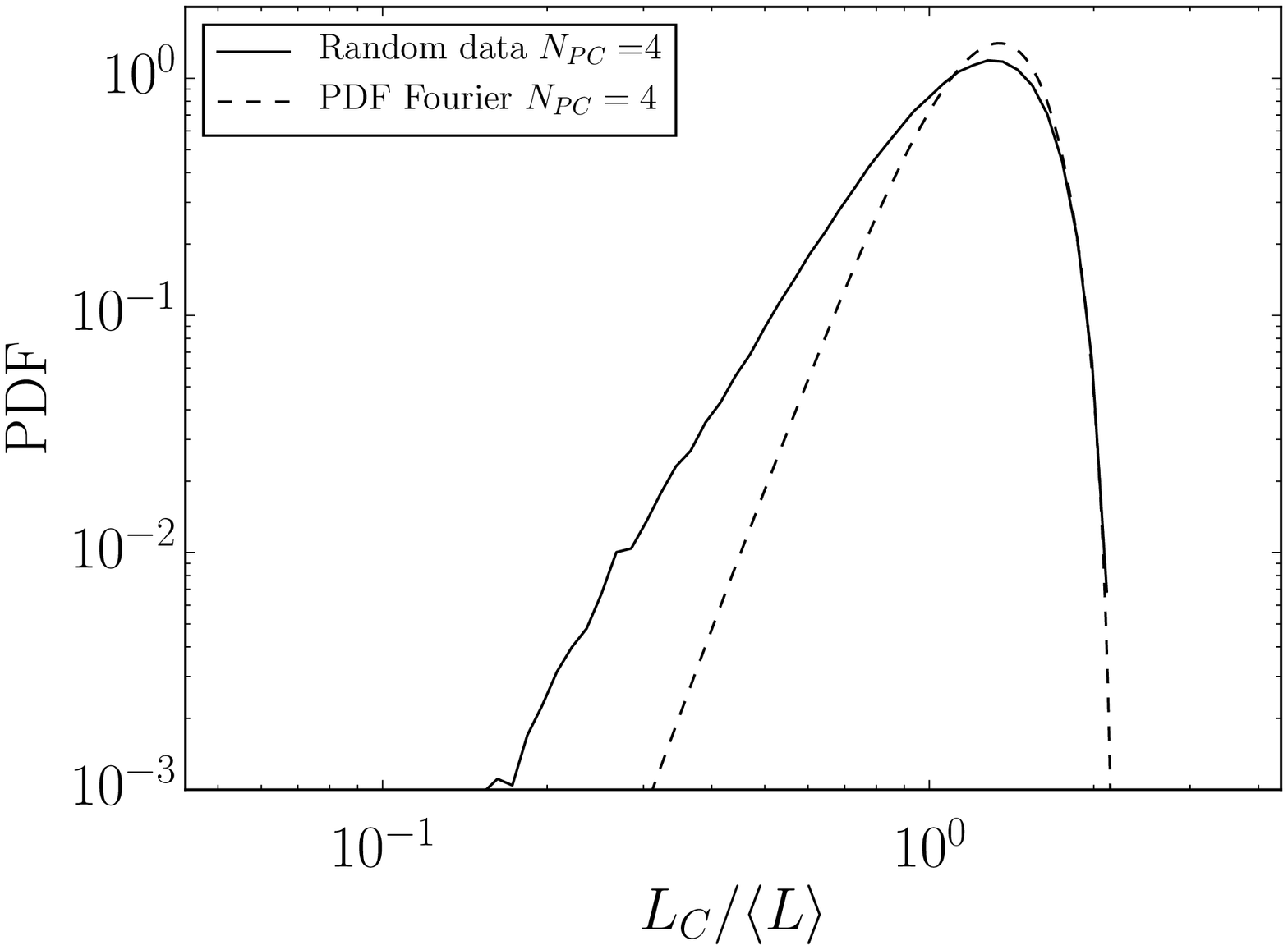}
		\caption{\label{cl-nn8}}
\end{subfigure}
\quad
\begin{subfigure}[h!]{0.48\textwidth}
		\includegraphics[scale=.4]{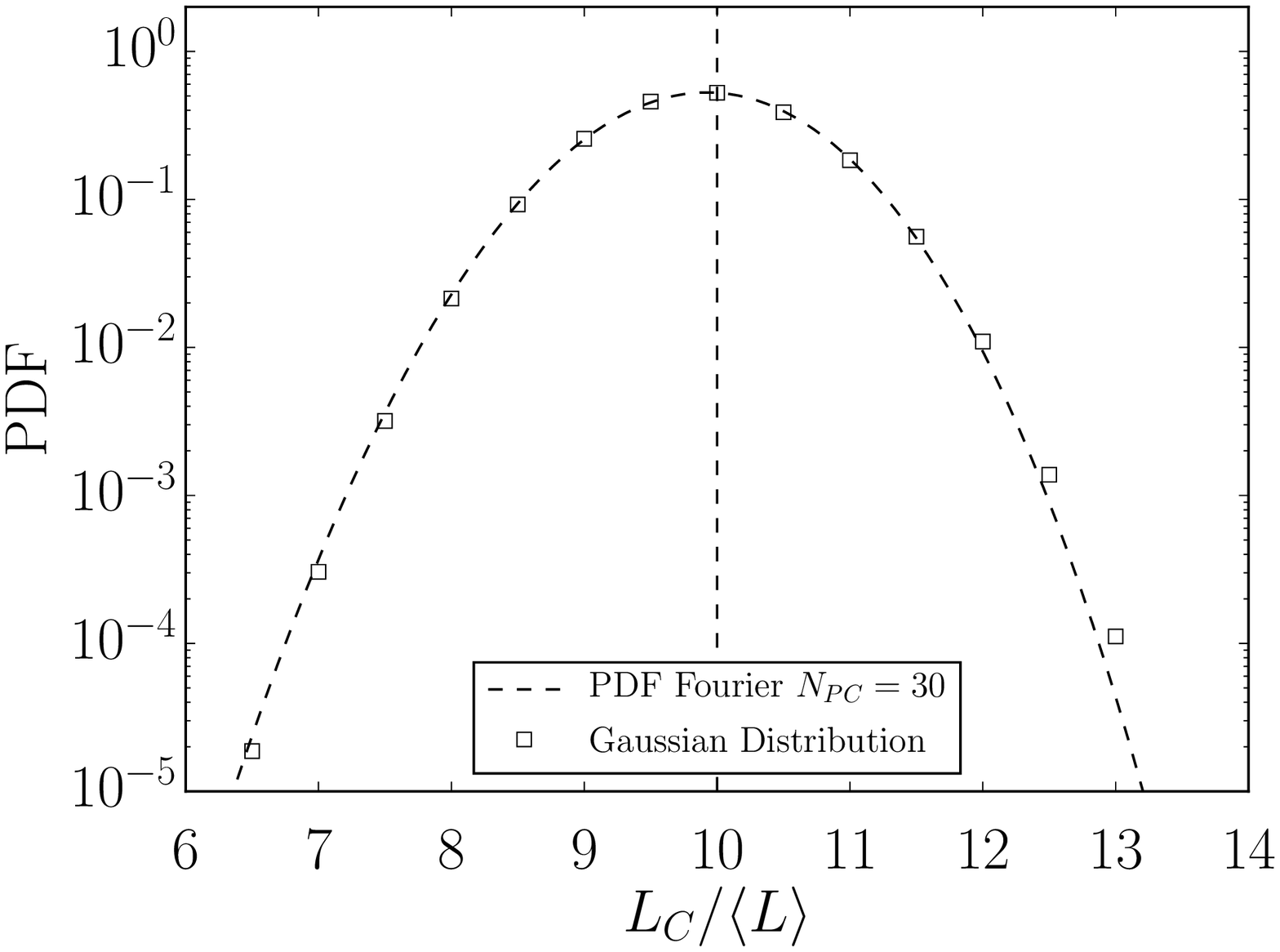}
		\caption{\label{cl-nn30}}

\end{subfigure}
\quad
\begin{subfigure}[h!]{0.48\textwidth}
		\includegraphics[scale=.4]{1D-Clusters-uptonn-MEAN.eps}
		\caption{\label{fig-clu-rdn25}}
\end{subfigure}

\caption{a) PDF of the cluster size $L_C/\langle L\rangle$, where $\langle L\rangle$ is the average Vorono\"{i} cell size.  The plot reveals that the model (equation \ref{eq:nn2}) has good agreement with the right tail of numerical generated data. The plot also shows the Fourier computed PDF. b) PDF of the cluster size $L_C/\langle L \rangle$.  The plot reveals that the model (equation \ref{eq:nn2}) has good agreement with the right tail of numerical generated data. The plot also shows the Fourier computed PDF. c)  PDF of the cluster size $L_C/\langle L \rangle$ for $N_{PC}=30$. This figure illustrates that as $N_{PC}$ increases the PDF tends to a normal distribution as expected by the central limit theorem \cite{hogg2005introduction}. d)  PDF of 1DVOA for normalized linear cluster size  $L_{C}/\langle L_{C} \rangle$ for a random uniform distribution, experimental data, and the model proposed. For all clusters having between  2 and 25 points ($N_{PC}$).  The model here proposed represents well the right tail of the RPP data, and the marker ($\square$) corresponds to data from \cite{Sumbekova2016a}, which computed the clusters with the condition $N_{PC}\geq 1$. The condition $\mathcal{V}\leq\mathcal{V}_{th}=0.55$ was employed for clustering computation.}
\end{figure}

For larger values of $N_{PC}$, the convolution in `physical' space becomes cumbersome, and thereby, the duality between the convolution and the Fourier transform, i.e., $\mathcal{F}\big\{F*G\big\}= \widehat{F}\cdot \widehat{G}$ is going to be employed to compute the PDF of $L_C/\langle L\rangle \vert_{N_{PC}}$, for $N_{PC}\geq3$, whence:

\begin{equation}
f_{{N_{PC}}}=\mathcal{F}^{-1}\big\{\prod_{k=1}^{N_{PC}}\widehat{F}\}
\label{eq:npcc}
\end{equation}

being $\widehat{F}$ the Fourier transform of the PDF found in equation \ref{eq:pdf}. This approach (see figures \ref{cl-nn2}-\ref{cl-nn8}) has good agreement regarding the PDF right tail despite the increase in statistical uncertainty due to the decrease in the number of samples for an increasing $N_{PC}$ (see figure \ref{hist-cl-2D-AA}) 

It is expected that as $N_{PC}\rightarrow\infty$ the PDF collapses into a normal/gaussian distribution due to the central limit theorem \cite{hogg2005introduction}, as seen in figure \ref{cl-nn30}. The average of this distribution is given by equation \ref{eq:NAVG}, e.g., for $N_{PC}=30$, $\langle L_C \rangle/\langle L\rangle \vert_{N_{PC}=30}\approx 10$.

\item The second assumption deals with the composition of the PDF for clusters having between $2\leq N_{PC}\leq N_{PC}^\star$, with $N_{PC}^\star=3,4,\ldots,N$ points.

The approach taken was to make a mixture model \cite{fruhwirth2006finite}, i.e., the composite PDFs will be a weighted sum of the \textbf{normalized PDFs} for each $N_{PC}$. This is written as:

\begin{align}
f_{2\leq N_{PC} \leq N}= \sum^N_{k=2} \alpha_{k-2} f_k \label{eq:cpdf}l\\
\sum^N_{k=2} \alpha_{k-2} = 1 \label{eq:we}\\
\int^{n\mathcal{V}_{th}}_0 f_n(\mathcal{Z}) d\mathcal{Z}=1 \label{eq:nom}
\end{align}

Being able to compute $f_k$, the weights $\alpha_{k-2}$ were modeled following a numerical experiment (see figure \ref{hist-cl-2D-AA}), which revealed that in 1D there is a predominantly presence of 2-point clusters, and that in turn this presence halves for 3-points clusters, and halves again for 4-point clusters, and so on. This can also be seen as the probability of finding a n-point cluster. The result is plotted in Figure \ref{fig-clu-rdn}, and it is repeated here for clusters between $2\leq N_{PC} \leq 25$ (see figure \ref{fig-clu-rdn25}),  the latter being the larger cluster detected in the numerical experiment. To change from  $L_C/\langle L \rangle$ to $L_C/\langle L_C\rangle$, it is straight forward following the chain rule \cite{hogg2005introduction};

\begin{equation}
f_Y(y)=f_X(v(y))\vert v^\prime (y)\vert
\end{equation}

which completes the explanation of the model here proposed. 

Given these previous results and the central limit theorem, one conjectures that for $L_C/\langle L \rangle\gg1$ the right tail of the clusters PDF will be within an envelope determined only by the weights $\alpha_i$ distribution, regardless of the presence of turbulence. Hence, it validates our model, as our interest is to examine right tail of such PDF.

The left tail however (as explained in the main text) behave as $f(\mathcal{M}_C/\langle \mathcal{M}\rangle< 1) \approx (\mathcal{M}_C/\langle \mathcal{M}\rangle)^\alpha$ behaves as twice this exponent, e.g., $\alpha \approx 2/5/8$ in 1D/2D/3D. The lower exponent than the respective on from our model $\alpha=3$ is due to our assumptions, as we  independently $N_{PC}$ cells PDFs, whereas in reality, due to the tessellation construction the distance between the cells `centers' is \textbf{correlated}, and such correlation seems to be particularly important for small, `compact' clusters, as for large clusters the joint PDF can successfully modeled as a product of marginal PDFs, as shown.

\end{itemize}

\subsection{Higher Dimensions}
\label{aa:hd}
\begin{figure}
	\centering
	\begin{subfigure}[h!]{0.48\textwidth}
			\includegraphics[scale=.4]{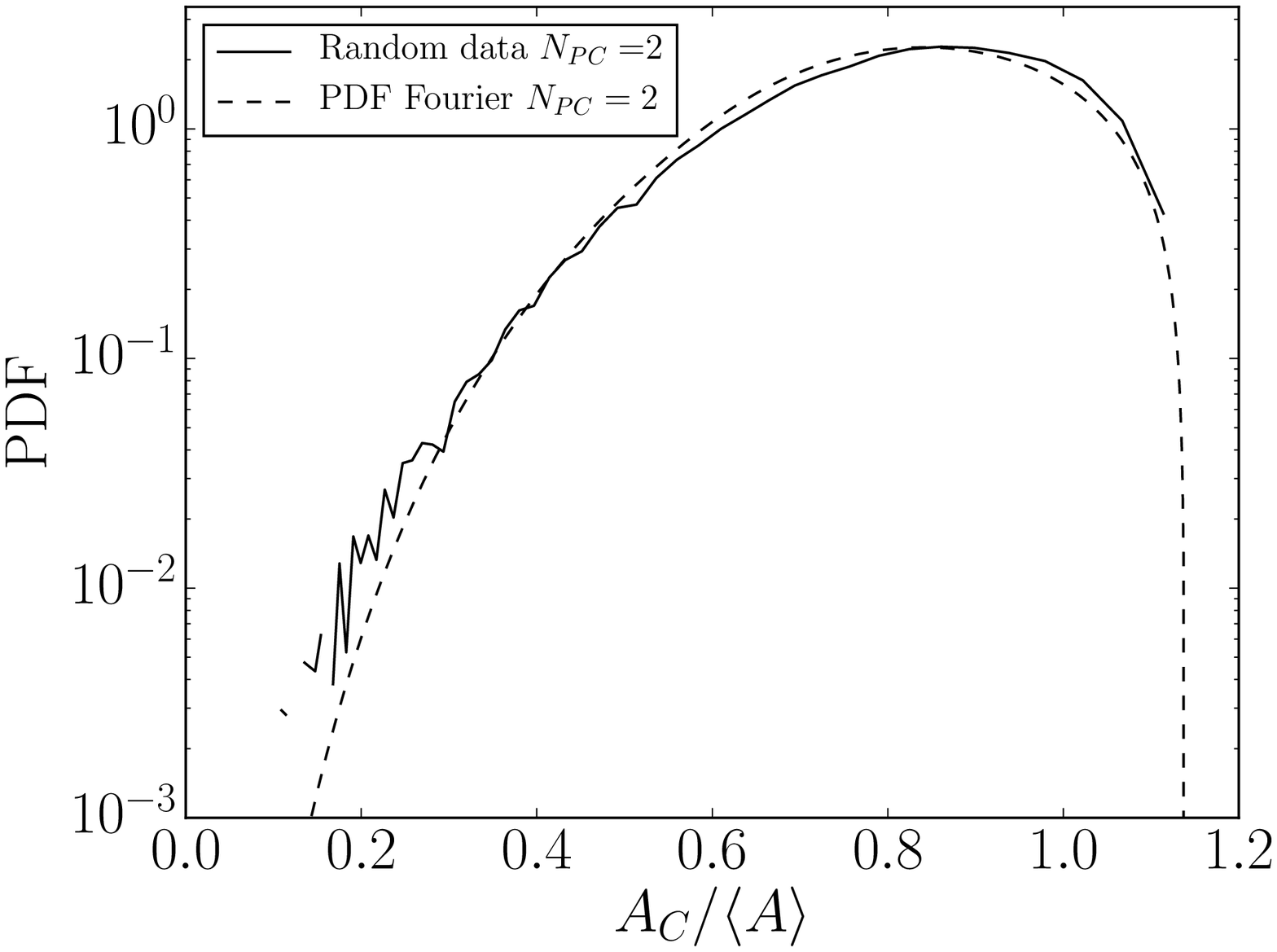}
			\caption{\label{cl-2d-nn2}}
	\end{subfigure}
\quad
\begin{subfigure}[h!]{0.48\textwidth}
		\includegraphics[scale=.4]{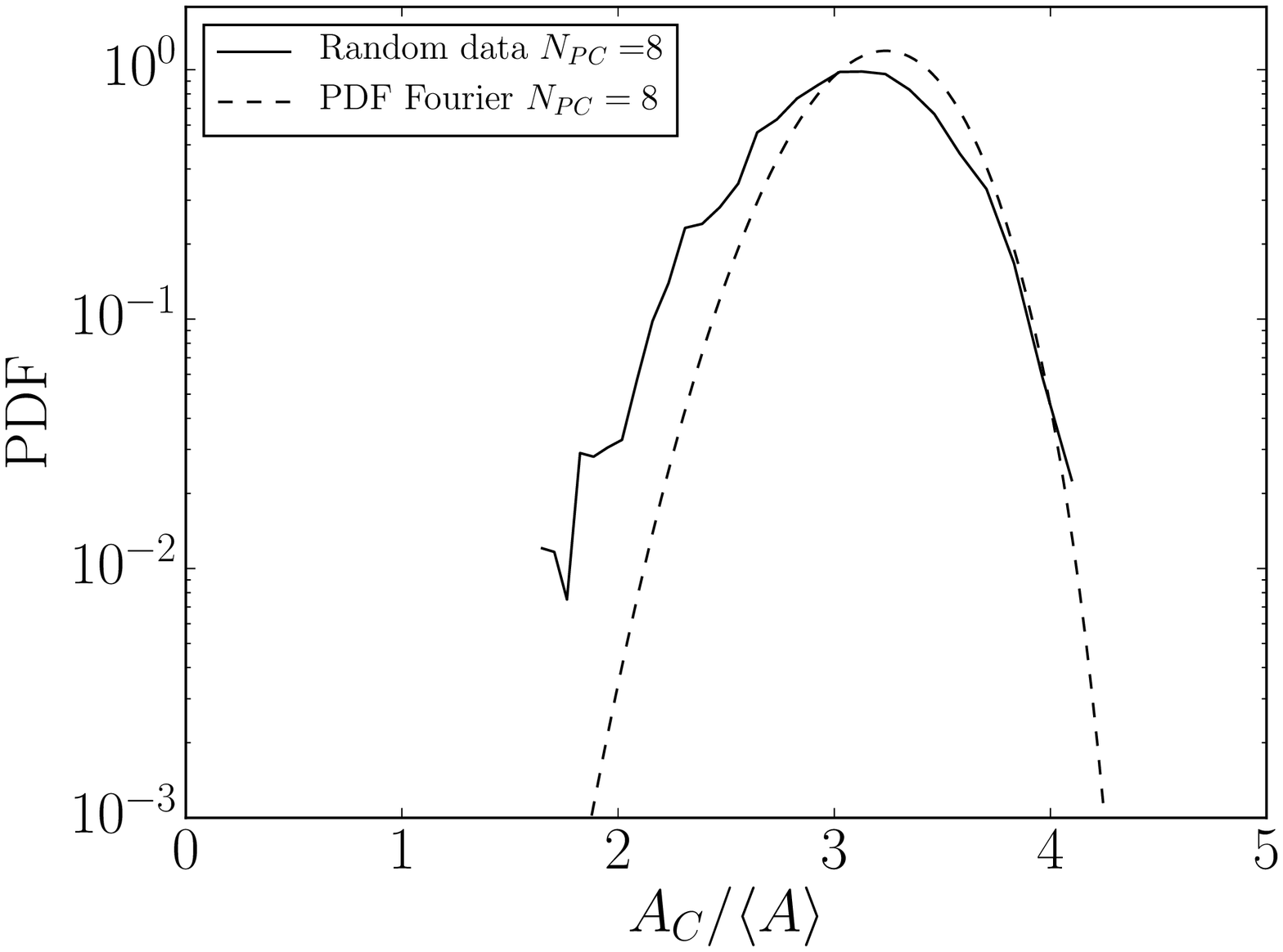}
		\caption{\label{cl-2d-nn8}}

\end{subfigure}

	\caption{a) PDF of the cluster size $A_C\langle A\rangle$.  The plot reveals that the model (equation \ref{eq:nn2}) has good agreement with the right tail of numerical generated data. The plot also shows the Fourier computed PDF. b) PDF of the cluster size $A_C$.  The plot reveals that the model (equation \ref{eq:nn2}) has good agreement with the right tail of numerical generated data. The plot also shows the Fourier computed PDF.}
\end{figure}

The extension to the previous arguments to higher dimensions has the pitfall of not having analytical PDFs available to do the computations, but rather fits proposed by Ferenc and N\'{e}da \cite{Ferenc2007}, and therefore, a mismatch is expected. The weights also have a different behavior than their counter part in 1D (see figure \ref{hist-cl-2D-AA}). For instance, in  2DVOA  the figures \ref{cl-2d-nn2}-\ref{cl-2d-nn8} reveal that there is an acceptable agreement again with the right tail of the numerical experiment. Proceeding into the compound PDF (equation \ref{eq:cpdf}), figure \ref{fig-clu-aa} shows that the model captures the changes in the behavior of the numerical PDF (inflection points), however the slope is not \textit{completely} well captured, and effect that might be due to not enough numerical samples, or the inaccuracy of the fits proposed by Ferenc and N\'{e}da \cite{Ferenc2007} (error propagation).

\bibliographystyle{plain}

\end{document}